\documentstyle[12pt,epsfig]{article}

\catcode`\@=11
\def\@citex[#1]#2{%
\if@filesw \immediate \write \@auxout {\string \citation {#2}}\fi
\@tempcntb\m@ne \let\@h@ld\relax \def\@citea{}%
\@cite{%
  \@for \@citeb:=#2\do {%
    \@ifundefined {b@\@citeb}%
      {\@h@ld\@citea\@tempcntb\m@ne{\bf ?}%
      \@warning {Citation `\@citeb ' on page \thepage \space undefined}}%
      {\@tempcnta\@tempcntb \advance\@tempcnta\@ne%
      \@tempcntb\number\csname b@\@citeb \endcsname \relax%
      \ifnum\@tempcnta=\@tempcntb 
        \ifx\@h@ld\relax%
          \edef \@h@ld{\@citea\csname b@\@citeb\endcsname}%
        \else%
          \edef\@h@ld{\ifmmode{-}\else--\fi\csname b@\@citeb\endcsname}%
        \fi%
      \else
        \@h@ld\@citea\csname b@\@citeb \endcsname%
        \let\@h@ld\relax%
      \fi}%
    \def\@citea{,\penalty\@highpenalty\,}%
  }\@h@ld
}{#1}}

\def\@citeb#1#2{{[#1]\if@tempswa , #2\fi}}
\def\@citeu#1#2{{$^{#1}$\if@tempswa , #2\fi }}
\def\@citep#1#2{{#1\if@tempswa , #2\fi}}

\def\bcites{         
        \catcode`\@=11
        \let\@cite=\@citeb
        \catcode`\@=12
}

\def\upcites{         
        \catcode`\@=11
        \let\@cite=\@citeu
        \catcode`\@=12
}

\def\plaincites{      
        \catcode`\@=11
        \let\@cite=\@citep
        \catcode`\@=12
}

\newcount\hour
\newcount\minute
\newtoks\amorpm
\hour=\time\divide\hour by 60
\minute=\time{\multiply\hour by 60 \global\advance\minute by-\hour}
\edef\standardtime{{\ifnum\hour<12 \global\amorpm={am}%
        \else\global\amorpm={pm}\advance\hour by-12 \fi
        \ifnum\hour=0 \hour=12 \fi
        \number\hour:\ifnum\minute<10 0\fi\number\minute\the\amorpm}}
\edef\militarytime{\number\hour:\ifnum\minute<10 0\fi\number\minute}

\def\draftlabel#1{{\@bsphack\if@filesw {\let\thepage\relax
   \xdef\@gtempa{\write\@auxout{\string
      \newlabel{#1}{{\@currentlabel}{\thepage}}}}}\@gtempa
   \if@nobreak \ifvmode\nobreak\fi\fi\fi\@esphack}
        \gdef\@eqnlabel{#1}}
\def\@eqnlabel{}
\def\@vacuum{}
\def\marginnote#1{}
\def\draftmarginnote#1{\marginpar{\raggedright\scriptsize\tt#1}}
\def\draft{
        \pagestyle{plain}
        \overfullrule=2pt
        \oddsidemargin -.5truein
        \def\@oddhead{\sl \phantom{\today\quad\militarytime} \hfil
        \smash{\Large\sl DRAFT} \hfil \today\quad\militarytime}
        \let\@evenhead\@oddhead
        \let\label=\draftlabel
        \let\marginnote=\draftmarginnote
        \def\ps@empty{\let\@mkboth\@gobbletwo
        \def\@oddfoot{\hfil \smash{\Large\sl DRAFT} \hfil}
        \let\@evenfoot\@oddhead}
        \def\@eqnnum{(\theequation)\rlap{\kern\marginparsep\tt\@eqnlabel}%
        \global\let\@eqnlabel\@vacuum}  }

\def\eqalign#1{\null\,\vcenter{\openup\jot\m@th
  \ialign{\strut\hfil$\displaystyle{##}$&$\displaystyle{{}##}$\hfil
      \crcr#1\crcr}}\,}
\def\eqalignno#1{\displ@y \tabskip\centering
  \halign to\displaywidth{\hfil$\@lign\displaystyle{##}$\tabskip\z@skip
    &$\@lign\displaystyle{{}##}$\hfil\tabskip\centering
    &\llap{$\@lign##$}\tabskip\z@skip\crcr
    #1\crcr}}

\def\section{\@startsection {section}{1}{\z@}{3.ex plus 1ex minus
 .2ex}{2.ex plus .2ex}{\large\bf}}
\def\subsection{\@startsection{subsection}{2}{\z@}{2.75ex plus 1ex minus
 .2ex}{1.5ex plus .2ex}{\bf}}        

\def\thefootnote{\arabic{footnote}}
\def\abstract{\if@twocolumn
\section*{Abstract}
\else 
\begin{center}
{\bf Abstract\vspace{-.5em}\vspace{0pt}}
\end{center}
\quotation
\fi}

\catcode`\@=12

\catcode`\@=11
\def\theequation{\arabic{equation}}
\catcode`\@=12

\bcites

\catcode`\@=11
\def\theequation{\thesection.\arabic{equation}}
\@addtoreset{equation}{section}
\@addtoreset{footnote}{section}
\@addtoreset{footnote}{subsection}
\catcode`\@=12

\newcommand{\beq}{\begin{equation}}
\newcommand{\beqa}{\begin{eqnarray}}
\newcommand{\bega}{\begin{array}}
\newcommand{\ea}{\end{array}}

\newcommand{\eeq}{\end{equation}}
\newcommand{\eeqa}{\end{eqnarray}}
\newcommand{\p}{\partial}
\newcommand{\N}{{\mathsf N}}

\newcommand{\Or}{{\cal O}}
\newcommand{\F}{{\bf F}}

\newcommand{\gss}{\gamma_{\sigma \sigma}}
\newcommand{\gst}{\gamma_{\sigma \tau}}
\newcommand{\intl}{\int_0^l d\sigma \,}

\newcommand{\lra}{\leftrightarrow}

\newcommand{\NN}{\mbox{$\cal N$}}
\newcommand{\NNN}{\mbox{$\cal N$}}

\newcommand{\tr}{\mbox{tr$\,$}}

\newcommand{\half}{\mbox{$1\over2$}}

\newcommand{\OO}{\mbox{$\cal O$}}
\newcommand{\PP}{\mbox{$\cal P$}}

\newcommand{\LL}{\mbox{$\cal L$}}

\newcommand{\e}{\epsilon}
\newcommand{\Ot}{{\tilde \Omega}}
\newcommand{\X}{{\cal X}}
\newcommand{\Y}{{\cal Y}}

\newcommand{\XII}[6]{\mbox{
\setlength{\unitlength}{0.7em}
\begin{picture}(9.5,3)
\put(0,1.5){\line(1,0){9}}
\put(0,-0.5){\line(1,0){9}}
\put(1,1.5){\line(1,-1){2}}
\put(3,1.5){\line(-1,-1){2}}
\put(5,1.5){\line(0,-1){2}}
\put(8,1.5){\line(0,-1){2}}
\put(5.5,0){\makebox (2,1){...}}
\put(1,0){\makebox (2,1){\tiny $\bullet$}}
\put(0.5,2){\makebox (1,1){$#1 \phantom{d\!\!\!}$}}
\put(0.5,-2){\makebox (1,1){$#2 \phantom{d\!\!\!}$}}
\put(2.5,2){\makebox (1,1){$#3 \phantom{d\!\!\!}$}}
\put(2.5,-2){\makebox (1,1){$#4 \phantom{d\!\!\!}$}}
\put(4.5,2){\makebox (1,1){$#5 \phantom{d\!\!\!}$}}
\put(4.5,-2){\makebox (1,1){$#5 \phantom{d\!\!\!}$}}
\put(7.5,2){\makebox (1,1){$#6 \phantom{d\!\!\!}$}}
\put(7.5,-2){\makebox (1,1){$#6 \phantom{d\!\!\!}$}}
\end{picture}$\phantom{_{\Big|}}$}}

\newcommand{\XIIm}[8]{\mbox{
\setlength{\unitlength}{0.7em}
\begin{picture}(9.5,3)
\put(0,1.5){\line(1,0){9}}
\put(0,-0.5){\line(1,0){9}}
\put(1,1.5){\line(1,-1){2}}
\put(3,1.5){\line(-1,-1){2}}
\put(5,1.5){\line(0,-1){2}}
\put(8,1.5){\line(0,-1){2}}
\put(5.5,0){\makebox (2,1){...}}
\put(1,0){\makebox (2,1){\tiny $\bullet$}}
\put(0.5,2){\makebox (1,1){$#1 \phantom{d\!\!\!}$}}
\put(0.5,-2){\makebox (1,1){$#2 \phantom{d\!\!\!}$}}
\put(2.5,2){\makebox (1,1){$#3 \phantom{d\!\!\!}$}}
\put(2.5,-2){\makebox (1,1){$#4 \phantom{d\!\!\!}$}}
\put(4.5,2){\makebox (1,1){$#5 \phantom{d\!\!\!}$}}
\put(4.5,-2){\makebox (1,1){$#6 \phantom{d\!\!\!}$}}
\put(7.5,2){\makebox (1,1){$#7 \phantom{d\!\!\!}$}}
\put(7.5,-2){\makebox (1,1){$#8 \phantom{d\!\!\!}$}}
\end{picture}$\phantom{_{\Big|}}$}}

\newcommand{\IIX}[8]{\mbox{
\setlength{\unitlength}{0.7em}
\begin{picture}(9.5,3)
\put(0,1.5){\line(1,0){9}}
\put(0,-0.5){\line(1,0){9}}
\put(1,1.5){\line(0,-1){2}}
\put(3,1.5){\line(0,-1){2}}
\put(6,1.5){\line(1,-1){2}}
\put(8,1.5){\line(-1,-1){2}}
\put(4.0,0){\makebox (2,1){...}}
\put(1,0){\makebox (12,1){\tiny $\bullet$}}
\put(0.5,2){\makebox (1,1){$#1 \phantom{d\!\!\!}$}}
\put(0.5,-2){\makebox (1,1){$#2 \phantom{d\!\!\!}$}}
\put(2.5,2){\makebox (1,1){$#3 \phantom{d\!\!\!}$}}
\put(2.5,-2){\makebox (1,1){$#4 \phantom{d\!\!\!}$}}
\put(5.5,2){\makebox (1,1){$#5 \phantom{d\!\!\!}$}}
\put(5.5,-2){\makebox (1,1){$#6 \phantom{d\!\!\!}$}}
\put(7.5,2){\makebox (1,1){$#7 \phantom{d\!\!\!}$}}
\put(7.5,-2){\makebox (1,1){$#8 \phantom{d\!\!\!}$}}
\end{picture}$\phantom{_{\Big|}}$}}

\newcommand{\III}[6]{\mbox{
\setlength{\unitlength}{0.7em}
\begin{picture}(9.5,3)
\put(0,1.5){\line(1,0){9}}
\put(0,-0.5){\line(1,0){9}}
\put(1,1.5){\line(0,-1){2}}
\put(4.5,1.5){\line(0,-1){2}}
\put(8,1.5){\line(0,-1){2}}
\put(5.0,0){\makebox (2,1){...}}
\put(1.5,0){\makebox (2,1){...}}
\put(0.5,2){\makebox (1,1){$#1 \phantom{d\!\!\!}$}}
\put(0.5,-2){\makebox (1,1){$#2 \phantom{d\!\!\!}$}}
\put(4.0,2){\makebox (1,1){$#3 \phantom{d\!\!\!}$}}
\put(4.0,-2){\makebox (1,1){$#4 \phantom{d\!\!\!}$}}
\put(7.5,2){\makebox (1,1){$#6 \phantom{d\!\!\!}$}}
\put(7.5,-2){\makebox (1,1){$#6 \phantom{d\!\!\!}$}}
\end{picture}$\phantom{_{\Big|}}$}}

\newcommand{\blobII}[6]{\mbox{
\setlength{\unitlength}{0.7em}
\begin{picture}(9.5,3)
\put(0,1.5){\line(1,0){9}}
\put(0,-0.5){\line(1,0){9}}
\put(1.5,1.5){\line(0,-1){2}}
\put(2.5,1.5){\line(0,-1){2}}
\put(5,1.5){\line(0,-1){2}}
\put(8,1.5){\line(0,-1){2}}
\put(5.5,0){\makebox (2,1){...}}
\put(1,0){\makebox (2,1)
{\large $\bullet \!\!\rule[0.04em]{0.7em}{0.4em} \!\!\bullet$}}
\put(0.5,2){\makebox (1,1){$#1 \phantom{d\!\!\!}$}}
\put(0.5,-2){\makebox (1,1){$#2 \phantom{d\!\!\!}$}}
\put(2.5,2){\makebox (1,1){$#3 \phantom{d\!\!\!}$}}
\put(2.5,-2){\makebox (1,1){$#4 \phantom{d\!\!\!}$}}
\put(4.5,2){\makebox (1,1){$#5 \phantom{d\!\!\!}$}}
\put(4.5,-2){\makebox (1,1){$#5 \phantom{d\!\!\!}$}}
\put(7.5,2){\makebox (1,1){$#6 \phantom{d\!\!\!}$}}
\put(7.5,-2){\makebox (1,1){$#6 \phantom{d\!\!\!}$}}
\end{picture}$\phantom{_{\Big|}}$}}

\begin{document}

\begin{titlepage}

\begin{center}
\hfill EFI-02-97\\
\hfill UCLA/02/TEP/18\\
\hfill hep-th/0208010\\

\vskip 2.5 cm
{\large \bf Strings in the near plane wave background and AdS/CFT}
\vskip 1 cm 
\renewcommand{\thefootnote}{\fnsymbol{footnote}}
{
Andrei Parnachev${}^1$ and Anton V. Ryzhov${}^2$ }

\setcounter{footnote}{0}

\vskip 0.5cm
${}^1$ {\sl Department of Physics and Enrico Fermi Institute,\\
University of Chicago, Chicago, IL 60637, USA\\ 
         {\tt andrei@theory.uchicago.edu}}

\vskip 0.3cm
${}^2$ {\sl Department of Physics and Astronomy, \\
            University of California, Los Angeles, LA, CA 90095-1547\\ 
         {\tt ryzhovav@physics.ucla.edu}}

\end{center}

\vskip 0.5 cm
\begin{abstract}
We study the AdS/CFT correspondence for string states which
flow into plane wave states in the Penrose limit.
Leading finite radius corrections to the string spectrum are compared
with scaling dimensions of finite R-charge BMN-like operators.
We find agreement between string and gauge theory results.

\end{abstract}

\end{titlepage}

\section{Introduction}
\label{section:intro}

The celebrated AdS/CFT correspondence asserts that 
the dual description of \NNN=4 four dimensional 
super Yang Mills is type IIB string theory 
in $AdS_5 \times S^5$ with self-dual
RR five-form field strength \cite{Maldacena:1997re,Gubser:1998bc,Witten:1998qj}. 
The radius of curvature of $AdS_5$ and $S^5$ 
scales like $R/l_s \sim (g_{\rm Y\!M}^2 N)^{1/4}\sim (g N)^{1/4}$. 
The spectrum of string states in this background 
corresponds to the spectrum of operators in SYM.
Part of the difficulty in directly verifying 
this proposal is that string quantization in the
presence of RR flux is notoriously difficult.
On the other hand type IIB supergravity, which
describes the dynamics of massless string modes, is
only valid for the large values of $R/l_s$, while
on the SYM side one can perform 
reliable computations only for small 't Hooft 
coupling $g N$.
Until recently, one mostly studied the properties of 
supergravity modes, and the corresponding protected SYM operators, 
appealing to nonrenormalization theorems to compare 
their correlators in the dual descriptions 
\cite{ADS-CFT:protected}. 

The GS superstring can be quantized exactly in the plane wave background 
\cite{Metsaev,Metsaev:2002re}, 
which can be viewed as the Penrose limit of the $AdS_5 \times S^5$ 
geometry \cite{BMN,Blau:2002dy}.
The limit involves scaling both the $AdS_5$ radius 
$R \rightarrow \infty$ and the R-charge $J \sim R^2$.
One considers 
states with finite plane wave light cone energy and momentum.
It has been proposed by Berenstein, Maldacena and Nastase (BMN)
\cite{BMN} that such string states correspond
to single trace operators in the gauge theory with certain phases
inserted.
Remarkably, the parameter controlling perturbative expansion of scaling
dimensions of such operators is $\lambda' = g N / J^2$, which
can be made small to allow reliable gauge theory computations.
BMN were able to resum the diagrams weighted by powers of $\lambda'$
and show precise agreement between the scaling dimensions of SYM operators 
and the light cone energies of corresponding string states. 
This has been further confirmed in \cite{Gross:2002su,Santambrogio:2002sb,Kristjansen:2002bb}.
The following development included studying string interactions both in 
the plane wave string theory and in the gauge theory
\cite{Kristjansen:2002bb,Spradlin:2002ar,Spradlin:2002rv,Klebanov:2002mp,Berenstein:2002sa,CFHMMPS,Kiem:2002xn,Huang:2002wf,Chu:2002pd,Lee:2002rm}. 

The plane wave limit is a dramatic improvement over 
being able to handle just the supergravity states and protected operators. 
But we would still like 
to get closer to the full AdS string theory.
One way to gain insight is to do systematic perturbation 
theory around the plane wave limit, taking $1/R^2$ 
as a small parameter. 
This approach has been tested in \cite{p1} on the
$AdS_3 \times S^3$ background with NS-NS flux.
String theory in this background is described by
an exactly solvable $SL(2) \times SU(2)$ WZNW model.
It has been shown \cite{p1} that one can recover
the exact string spectrum at small coupling $g$ to the next to
leading order in $1/R^2$ expansion.

In the present paper we use this approach to determine
the leading order finite radius corrections to the string spectrum
in $AdS_5 \times S^5$.
On the Yang Mills side, the corresponding calculation involves refining the 
definition of BMN operators and computing their scaling dimensions.
We work at 
small string coupling 
$g$, which corresponds
to computing only planar diagrams in the gauge theory.
Furthermore, we consider only the leading non-trivial term in 
the $\lambda'$ expansion. 
The calculation of scaling dimensions in SYM then reduces to
computing the matrix of two-point functions and its subsequent
diagonalization. 
We identify the gauge theory operator 
which corresponds to the light cone worldsheet Hamiltonian, 
and show that its matrix elements relevant for 
diagonalization agree with the string theory results.
Hence we conclude that to the accuracy we are working at,
the scaling dimensions of gauge theory operators 
agree with the spectrum of string states in $AdS_5 \times S^5$.

The paper is organized in the following way. 
In section \ref{section:worldsheet} 
we describe how to quantize the 
string in the background which includes 
the $\OO(1/R^2)$ corrections to the plane wave metric, 
and show how to compute the leading corrections to the 
spectrum of bosonic plane wave states.
In section \ref{section:SYM} we explain how the definition of BMN 
operators should be extended to include finite $J$ effects. 
There we also establish agreement 
between string and SYM results for a subset of 
matrix elements of the light cone Hamiltonian. 
In section \ref{section:discussion} we discuss our results
and mention possible future developments. 
In appendix \ref{section:gsw way} we present
an alternative technique, based on the formalism of \cite{GSW},
for computing $1/R^2$ corrections in string theory.
The results for  physical quantities are the same
as in section \ref{section:worldsheet}.
Appendix \ref{section:feynman rules} contains 
the tools we use in the SYM calculations. 
In appendix \ref{section:equalitygeneric} we generalize
the results of section \ref{section:SYM}.

{\bf Note added:} 
As we were completing this paper, we found out that 
related issues are addressed in \cite{Frolov:2002av,Schwarz,Tseytlin}.

\section{Corrections to the plane wave string spectrum}
\label{section:worldsheet}

In this section, we do perturbation theory 
on the worldsheet following the method described in \cite{p1}.
We start by outlining the procedure used in 
\cite{Metsaev,Metsaev:2002re,BMN} for deriving the 
leading order spectrum in the Penrose limit of $AdS_5 \times S^5$.
The $AdS_5 \times S^5$ metric is 
\begin{eqnarray}
\label{eq:exact metric}
ds^2 = R^2 
\left[
- dt^2 \cosh^2 \!\! \rho + d\rho^2 + \sinh^2 \!\! \rho \, d\Omega_3^2 
+ d\psi^2 \cos^2 \! \theta + d\theta^2 + \sin^2 \! \theta \, d\Omega_3'\!{}^2 
\right].
\end{eqnarray}
The Penrose limit of this geometry is obtained by zooming in 
on the neighborhood of a lightlike geodesic circling the equator of $S^5$. 
This is done by changing variables as 
\begin{eqnarray}
\label{eq:penrose variables:def}
X^+ = {1\over2} (t + \psi)
,\quad
X^- = {1\over2} (t - \psi) R^2 
,\quad
\rho = {r \over R}
,\quad
\theta = {y \over R},
\end{eqnarray}
and taking $R$ to be large, while keeping 
$|X^\pm|, r, y$ finite. 
At leading order in $1/R^2$, 
the $AdS_5 \times S^5$ metric (\ref{eq:exact metric}) reads 
\begin{eqnarray}
\label{eq:metric:leading order}
ds^2_0 = - 4 dX^- dX^+ - (r^2+y^2) dX^+ dX^+ + dr_i \, dr_i + dy_i \, dy_i.
\end{eqnarray}
Coordinates $y_i$ and $r_i$ parameterize two copies of $R^4$, 
but the $SO(8)$ symmetry of the metric (\ref{eq:metric:leading order}) 
is broken down to $SO(4) \times SO(4)$ by the
RR flux
\begin{eqnarray}
\label{eq:F-field:leading order}
F_{+1234} = F_{+5678} = \mbox{const}.
\end{eqnarray}

We would like to quantize type IIB superstring in the background 
(\ref{eq:metric:leading order}), (\ref{eq:F-field:leading order}). 
As was shown in \cite{Metsaev,Metsaev:2002re}, the way to do this is to 
look at the sigma-model part of the GS action, 
and use $\kappa$-symmetry in light-cone gauge 
to determine the rest of the worldsheet action. 
Bosons and fermions decouple for the plane wave background 
(\ref{eq:metric:leading order}) in light-cone gauge 
\cite{Metsaev,Metsaev:2002re}.
We will only be interested in the bosonic part 
of the full superstring action. 
The light cone gauge for bosonic fields is specified by
\beqa
\label{eq:lc:polch:start}
  X^+&=&\tau, \\ \nonumber
  \p_\sigma \gss&=&0, \\ \nonumber
  {\rm det} \gamma_{\alpha \beta}&=&-1, 
\eeqa 
where the worldsheet coordinates are 
$\tau \in (-\infty, \infty)$, $\sigma \in [0,l]$.
The worldsheet metric can be written as \cite{Polch}
\beq
  \gamma^{\alpha \beta}=\left(  \begin{array}{cc}
   -\gss(\tau) & \gst(\tau, \sigma)  \\
   \gst(\tau, \sigma) &  \gss^{-1}(\tau) (1-\gst^2(\tau, \sigma))
   \end{array} \right). 
\eeq

In this section we consider only the $y$ part of the theory.
The $r$ part can be included by noticing
that (\ref{eq:metric:leading order}) and (\ref{eq:F-field:leading order})
are invariant under $y \lra r$ while in the $\Or(1/R^2)$
correction to the plane wave metric $y$ and $r$ terms come
with opposite signs (see below).
This means that to restore the $r$ terms in the final result
one needs to copy the $y$ part, substitute $y \rightarrow r$ 
and flip the sign in front of the $\Or(1/R^2)$ terms.
This is confirmed in appendix \ref{section:gsw way}, where
explicit calculations are performed.

In the light cone gauge (\ref{eq:lc:polch:start}) 
the bosonic part of the Lagrangian is 
\beq
\label{l00}
L_0{=}{-}{1 \over 4 \pi} \int_0^l \left\{ \gss \bigg[
          4 {\dot X}^- {+}\sum_i ( y_i y_i{-} {\dot y}_i {\dot y}_i ) \bigg] 
{-} 2 \gamma_{\sigma\tau} 
\bigg[ 2 (X^-)' {-} \sum_i\dot y_i y_i' \bigg]
          {+}\gss^{-1} (1{-}\gamma_{\sigma\tau}^2) \sum_i y_i' y_i' \right\},
\eeq
where we used the leading order spacetime metric (\ref{eq:metric:leading order}). 
The equation of motion for the worldsheet metric (Virasoro constraints)
are
\beq
\label{ymi}
 (X^-)'={1 \over 2} \sum_i \dot y_i y_i'
, \qquad 
        {\dot X}^- = {1 \over 4} \sum_i
\left[ \dot y^i\dot y^i + y_i' y_i' - y_i y_i\right].
\eeq
One can use the equation of motion for 
$X^-$ and the leftover gauge freedom $\sigma \to \sigma + f(\tau)$ 
to set $\gamma_{\sigma\tau}=0$ in (\ref{l00}) \cite{Polch}. 
The equation of motion for the zero mode of $X^-$
implies that $\gss$ is related to the conserved light cone momentum 
$\PP_- = - i {\partial \over \partial X^-}$. 
Choosing the gauge 
\beq 
\label{gc}
l=2 \pi \eta,
\quad\mbox{where $\eta \equiv - {1\over2} \PP_-$},
\eeq
sets $\gss=1$ at the leading order in $1/R^2$.
The plane wave Hamiltonian that follows from  (\ref{l00}) can therefore be written as
\beq
\label{h0}
H_0={1 \over 4 \pi} \intl \sum_i \left[ 
           (2 \pi)^2 P^i_y P^i_y + y_i y_i  +y_i' y_i' \right].
\eeq 
where $P_y^i=\dot y_i/2 \pi$. 
The worldsheet theory of a light cone string is massive 
in the plane wave background.
The fields can be expressed in terms of eigenmodes
\beq
\label{xi}
  y_i={i \over \sqrt{2}} \sum_{n} {1 \over \sqrt{w_n}}
    \left[y^i_n-y^i_{n}{}^\dagger \right],
\eeq
where the  $\tau,\sigma$-dependent oscillators $y^i, y^i_n{}^\dagger$ are defined as 
\beq
  y^i_{n}=\alpha^i_{n} e^{-i w_n \tau -i n \sigma \over \eta},   \qquad
      y^i_{n}{}^\dagger=\alpha^i_{n}{}^\dagger e^{i w_n \tau +i n \sigma \over \eta}, \\ 
\eeq
and the frequencies are given by 
\beq
  w_n=\sqrt{\eta^2+n^2}.
\eeq
Substituting the field expansions into (\ref{h0}) diagonalizes
the plane wave Hamiltonian
\beqa
\label{h0o}
 H_0={1 \over \eta} \sum_{i,n} w_n N^i_{n},    
\eeqa
where 
$N^i_{n}= y^i_n{}^\dagger y^i_{n}$. 
The normal ordering constant cancels between bosons and fermions 
by virtue of spacetime supersymmetry, 
so we do not include it in (\ref{h0o}).
The leading terms in the expansion of $H_0$ in powers of $1/\eta^2$
are
\beq
\label{h0e}
   H_0=\sum_{i,n} N^i_n + {1 \over 2 \eta^2} \sum_{i,n} n^2 N^i_n+\Or\left({1 \over \eta^4}\right).
\eeq
In addition, we have the level matching condition 
\beq
\label{eq:level matching:w/s}
   \sum_{i,n} n N^i_n = 0.
\eeq

To compute $\OO(1/R^2)$ corrections 
to the string spectrum in the plane wave background, 
one would add the $\Or(1/R^2)$ correction $ds_1^2$ to 
the leading metric $ds_0^2$, write down the bosonic
part of the light cone Lagrangian, 
and then use $\kappa$-symmetry
to write the full GS action.
Subsequently the system can be quantized perturbatively in $1/R^2$. 
%
Expanding (\ref{eq:exact metric}) to next to leading order in $1/R^2$ we have
\begin{eqnarray}
\label{eq:metric:correction}
ds_1^2 = 
{1\over R^2} 
\left[
- 2 dX^- dX^+ (r^2-y^2) - {1\over3} (r^4-y^4) dX^+ dX^+ + 
{1\over3} (r^4 d\Omega_3{}^2 - y^4 d\Omega'_3{}^2) 
\right].
\end{eqnarray}
%
The bosonic part of the $\Or(1/R^2)$ Lagrangian is therefore
quartic in the fields.
The leading form of the $\kappa$-symmetry then implies that
the fermionic part of the $\Or(1/R^2)$ GS action is at most 
bi-quadratic in bosons and fermions.
We are considering corrections to the spectrum of bosonic states, 
so the fermionic part of the action can only contribute
diagonal matrix elements of the type
\beq
\label{not}
   {1 \over R^2} \sum_{i,n} f(w_n) N^i_n,
\eeq
where $f(w_n)$ is some function.
Fixing the exact form of $f(w_n)$ in (\ref{not}) requires dealing with
the $\Or(1/R^2)$ fermionic part of the superstring  action.
This we have not bothered to do.
We also drop all terms that are due to the normal ordering
of bosonic operators in all subsequent calculations.

Using the identities
$dy_i dy_i = dy^2+y^2 d\Omega^2$ and $y dy = y_i dy_i$ 
we can write 
$y^4  d\Omega_3'{}^2 = y_i y_i dy_j dy_j - y_i y_j dy_i dy_j$
and deduce the correction to the leading order Lagrangian (\ref{l00})
\beqa
\label{l1}
  L_1&=&{1 \over 4 \pi R^2} \intl \Bigg[ 
   {1 \over 3} \sum_i y_i^4
          {-} {1 \over 3} \sum_{i \neq j} \left[ y_i^2 ({\dot y}_j^2{-}y_j^2{-}(y_j')^2)
                {+}y_i y_j (y_i' y_j'{-}{\dot y}_i {\dot y}_j) \right]  \\ \nonumber 
   &&\qquad \qquad +{1 \over 2} y^2 {\dot X}^- \Bigg].
\eeqa
Terms proportional to $\gst$ are higher order in $1/R^2$ 
and do not contribute to (\ref{l1}). 
As explained in \cite{p1}, for the purpose of computing the
leading corrections to the spectrum, the correction to the Hamiltonian
equals minus the correction to the Lagrangian.%
\footnote{
	One can convince oneself that this is the case by
	perturbing the Lagrangian, 
	calculating the canonically conjugate momenta, 
	and keeping only terms up to $\Or(1/R^2)$ in the Hamiltonian. 
	In \cite{p1} the zero mode of $X^-$ was treated separately, 
	but one can show that this is not necessary.
	} 
The correction to the plane wave Hamiltonian can
therefore be written as
\beqa
\label{h1:1 plus 2}
  H_1&=& {1 \over 4 \pi R^2} \intl \Bigg[ 
   -{1 \over 3} \sum_i y_i^4
          {+} {1 \over 3} \sum_{i \neq j} \bigg[ y_i^2 
  [ (2 \pi P^j_y)^2{-}y_j^2{-}(y_j')^2] \\ \nonumber && \quad
                {+}y_i y_j (y_i' y_j'{-}(2 \pi)^2 P^i_y P^j_y       ) \bigg]  
       - {1 \over 2} \sum_{i,j} y_i^2 [ 
                      (2 \pi P^j_y)^2+(y_j')^2-y_j^2 ] \Bigg],
\eeqa
where in rewriting the last term we used the Virasoro constraint 
[the second equation in (\ref{ymi})]. 

Next we expand (\ref{h1:1 plus 2}) in modes (\ref{xi}). 
We are interested in first order corrections 
to the energies, so we only need to compute matrix elements 
of $H_1$ between degenerate states. 
Plane wave string states are 
\beq
\label{pwstates}
 y_{n_1}^{i_1}{}^\dagger  \ldots  y_{n_k}^{i_k}{}^\dagger  \ldots |\eta \rangle.
\eeq
They 
are degenerate only when the two sets of worldsheet momenta 
$(n_1,\ldots n_k,\ldots)$ and $(n_1',\ldots n_k',\ldots)$ 
are permutations of one another. 
Thus the only relevant terms in $H_1$ 
are of 
the form $y_k y^\dagger_k y_l y^\dagger_l$. 
Diagonal contributions come from 
$y_k^i y_k^i{}^\dagger y^j_l y_l^j{}^\dagger$;
they add up to 
\beq
\label{eq:adrei:diag}
 H_1^D = {1 \over 2 \eta R^2} \left( {1 \over 2} \sum_{i;n} {n^2 (N^i_n)^2 \over w_n^2}
        -\sum_{i,j;m,n} {n^2 N^i_n N^j_m \over w_m w_n}          \right).
\eeq
The relevant off-diagonal terms are of the form
$y^i_m{}^\dagger y^i_n y^j_n{}^\dagger y^j_m$, $i \neq j$, $m \neq n$; 
and
$y^i_m{}^\dagger y^i_n{}^\dagger y^j_m y^j_n$, $i \neq j$.
These add up to 
\beq
\label{eq:adrei:off-diag}
  H_1^{OD}={1 \over 2 \eta R^2 } \, \sum_{i \neq j;\,m \neq n}
  {n m \over w_n w_m}     
     ( y^i_m{}^\dagger y^i_n{}^\dagger y^j_m y^j_n  -y^i_m{}^\dagger y^i_n y^j_n{}^\dagger y^j_m)+
 {1 \over 4 \eta R^2 }  \sum_{i \neq j; n} {n^2 \over w_n^2}     
      y^i_n{}^\dagger y^i_n{}^\dagger y^j_n y^j_n.
\eeq     
Expanding (\ref{eq:adrei:diag}) and (\ref{eq:adrei:off-diag})
in powers of  $1/\eta$ we obtain
\beq
\label{h1diag}
    H_1^D = {1 \over 2 \eta^3 R^2} \left({1 \over 2} \sum_{i;n} n^2 (N^i_n)^2 
        -\sum_{i,j;m,n} n^2 N^i_n N^j_m          \right) +\Or\left({1 \over \eta^5 R^2}\right)
\eeq
and
\beqa
\label{h1offdiag}
  H_1^{OD}&=&{1 \over 2 \eta^3 R^2 } \, \sum_{i \neq j;\,m \neq n}
  n m      
     ( y^i_m{}^\dagger y^i_n{}^\dagger y^j_m y^j_n  {-}y^i_m{}^\dagger y^i_n y^j_n{}^\dagger y^j_m)
 \\ \nonumber && \quad
+{1 \over 4 \eta^3 R^2 }  \sum_{i \neq j; n} n^2 
      y^i_n{}^\dagger y^i_n{}^\dagger y^j_n y^j_n  
      {+}\Or\left({1 \over \eta^5 R^2}\right),
\eeqa  
respectively.
The leading $1/\eta$ term in $H_1$ is a sum of these two expressions.

An alternative derivation is given in appendix \ref{section:gsw way}, 
where more details are provided.


%
%
%
%

\section{Anomalous dimensions and AdS/CFT}
\label{section:SYM}

We now turn to 
the boundary $\NN=4$ Super Yang-Mills theory.
Our starting point will be the BMN operators \cite{BMN} which
correspond to plane wave states 
in the Penrose limit. 
One can still regard plane wave states as 
belonging to 
the Hilbert
space of the full $AdS_5 \times S^5$ theory, even though they are no longer
eigenstates of the full Hamiltonian.
As explained in the previous section, departing from the Penrose
limit corresponds to turning on perturbative corrections to the
plane wave Hamiltonian.
Eigenstates of the full Hamiltonian can 
be found using ordinary 
quantum-mechanical perturbation theory.

SYM operators which correspond to string eigenstates
must have definite conformal dimensions.
Such operators may be obtained from a complete set of operators 
by diagonalizing the matrix of their two-point functions.
This procedure is 
analogous to the
diagonalization of the string theory Hamiltonian.
We find that 
the spectra computed on both sides of the correspondence
match, and the operator defined by the
matrix of two-point functions is the SYM counterpart 
of the string Hamiltonian.

This section is organized as follows.
In section \ref{section:operators} we define operators that 
correspond to plane wave 
states away from the strict Penrose limit.
In section \ref{section:anomalous dims} we show how the matrix of
two-point functions is related to the string Hamiltonian.
In section \ref{section:equalitysimple} 
we match the 
matrix elements
of the light cone Hamiltonian between the string and the gauge theory.
We analyze a simple case where all of the excited modes 
have distinct $SO(4)$ indices and none of them 
is excited more than once.
The most general case is treated in appendix \ref{section:equalitygeneric}.
Feynman rules are discussed in appendix \ref{section:feynman rules}.

\subsection{Operators}
\label{section:operators}
The important assumption that we start with is that 
suitably refined BMN operators continue to correspond to
plane wave states, regarded as states in the Hilbert space
of $AdS_5 \times S^5$,  even away from the plane wave limit.
To define the right operators we will follow closely the
logic of BMN.
We start with the operator which corresponds to the light cone vacuum
\beq
\label{ovac}
   {1 \over \sqrt{\Omega} } \tr[ z^J] 
\quad \lra \quad  |\eta  \rangle,
\eeq
where $z={1\over\sqrt2} (\phi^5+i \phi^6)$ and $\Omega$
is a normalization constant 
(more about this below).
For the ground state (\ref{ovac}) there is a relation
$J=R^2 \eta$, but this gets modified by
$\Or(1/R^2)$ terms for excited states.

SYM operators which correspond to 
states with excited zero modes can be generated 
by acting on the light cone ground state (\ref{ovac}) 
with generators of the global symmetry group. 
The generators that we will be interested in are rotations in $ij$ plane,
denoted by $T_{ij}$ and their combinations 
$T_{i z}= {1\over \sqrt2} (T_{i5}+ i T_{i6})$
and 
$T_{i\bar z}={1\over \sqrt2} (T_{i5}- i T_{i6})$.
They act on the fields as
\beqa
  [T_{iz},z]&=&0, \qquad [T_{iz},{\bar z}]=\phi^i, \qquad 
[T_{iz}, \phi^j] = -z \, \delta_i^j, \\ \nonumber
  [T_{i\bar z},z]&=&\phi^i, \qquad  [T_{i\bar z},\bar z]=0, \qquad
  [T_{i\bar z}, \phi^j] = -{\bar z} \, \delta_i^j.
\eeqa
On the worldsheet we have a correspondence
\beq
   T_{iz} \lra  y_0^i, \quad  T_{i\bar z} \lra  y_0^i{}^\dagger.
\eeq
Consider as an example the operator corresponding to
the state $ y_0^i{}^\dagger y_0^j{}^\dagger |\eta \rangle, \, i \neq j$.
It is obtained by computing successive commutators of 
$T_{i\bar z}$ and  $T_{j\bar z}$ with (\ref{ovac}).
Either of these generators can turn any $z$ in the string of $z$-s into
$\phi^i$ or $\phi^j$ respectively.
The result is therefore the sum of 
$\tr[ z,\,\ldots \phi^i\, z\, \ldots \phi^j\, z \ldots]$
over all possible positions of inserted $\phi$'s:
\beq
\label{bps}
  {1 \over \sqrt{\Omega} } \left[ \sum_{a=0}^J \sum_{b=a}^J
    \tr[ z^a\, \phi^i\,z^{b-a}\, \phi^j\, z^{J-b}] +(i \lra j) \right]
\quad \lra \quad 
        y_0^i{}^\dagger y_0^j{}^\dagger |\eta \rangle.
\eeq
This formula has an obvious generalization for
higher number of $\phi$ insertions, as long as no label 
appears more than once.
If some of the $\phi$'s indices do coincide, 
$T_{i\bar z}$ can act on the same field. 
In this case, $z$ is first turned into $\phi^i$, 
and then into $-\bar z$.
For example, in the case of two $\phi$ insertions we have
\beq
\label{bps2}
  {1 \over \sqrt{\Omega} } \left( 2 \sum_{a=0}^J \sum_{b=a}^J
    \tr[ z^a\, \phi^i\,z^{b-a}\, \phi^i\, z^{J-b}] - \sum_{a=0}^{J+1}
               \tr[ z^a\, {\bar z}\,z^{J+1-a}] \right)
\quad \lra \quad 
        y_0^i{}^\dagger y_0^i{}^\dagger |\eta \rangle.
\eeq
To construct an operator with 
three $\phi$'s with the same index inserted,
one should act by  $T_{i\bar z}$ on both terms in (\ref{bps2})
to produce
\beq
\label{bps3}
  {1 \over \sqrt{\Omega} } \left( \sum
    \tr[ z \ldots \phi^i \ldots \phi^i \ldots \phi^i\ldots] - 3 \sum
               \tr[ z\ldots \phi^i\ldots {\bar z}\ldots] \right)
\quad \lra \quad 
        y_0^i{}^\dagger y_0^i{}^\dagger y_0^i{}^\dagger |\eta \rangle,
\eeq
where dots stand for a bunch of $z$'s and the sum is over all
possible positions of the insertions.
The second sum in (\ref{bps3}) has $J+1$ times
fewer terms than the first sum, and is subleading when it comes to
computing two-point functions.
Throughout this paper we are interested in the subleading
corrections in $1/J \sim 1/\eta R^2$, and therefore 
we should keep this term.
If we act with $T_{i\bar z}$ one more time, 
a term 
$3 \sum \tr[ z\ldots {\bar z}\ldots {\bar z}\ldots]$ 
appears when $T_{i\bar z}$ hits the $\phi^i$ 
in the second sum in (\ref{bps3}).
This piece is $\Or(1/J^2)$ compared to the leading term, 
so we can drop it. 

In general, when an arbitrary number of zero modes
excited, the corresponding SYM operator is
\beqa
\label{genericbps}
 \Or&=& {\tilde \Or} - \Or_*, \\ 
 {\tilde \Or}
     &=& {1 \over \sqrt{\Omega} }  \sum  \tr[ z \ldots \phi^{i_1} \ldots \phi^{i_k} \ldots ], \\
 \Or_*&=& {1 \over \sqrt{\Omega} }
         \sum_{(p,q):  \,i_p=i_q}
         \tr[ z\ldots \phi^{i_1} \ldots \phi^{i_k} \ldots
               \check{\phi}^{i_p} \ldots  \check{\phi}^{i_q} 
                         \ldots {\bar z}\ldots],
\eeqa
where $\check{\phi}^{i_p}$ stands for ${\phi}^{i_p}$ being omitted from
the string of operators and the sum in $\Or_*$ runs over
all possible pairs of  $({\phi}^{i_p},{\phi}^{i_q}) $ with the same indices.
When writing 
(\ref{genericbps}), 
we omitted terms which appear when $T_{i\bar z}$ hits
the same field more than twice, as such are $\Or(1/J^2)$.
When all $\phi$'s inserted have different flavors, the
operator $\Or_*$ vanishes and we have $\Or= {\tilde \Or}$.

Next we turn to the construction of operators which correspond to 
general 
string states (\ref{pwstates}). 
Such operators must satisfy a few  necessary requirements.
First, if only the zero modes are excited,
they must reduce to the BPS operators described. 
Second, they must vanish unless the level matching condition 
\beq
\label{lm}
\sum_{i,n} n N^i_{n}=0,
\eeq
is satisfied.
Finally, our operators must reduce to the BMN operators 
as $J \rightarrow \infty$.

Suppose there is a total of $\N$ oscillators excited, 
\beq
   \N=\sum_i \N_i, \qquad \N_i=\sum_{n}  N^i_n.
\eeq
Due to the cyclicity of the trace,
\beq
\label{co}
  {1 \over \sqrt{\Omega} } 
    \tr[ z\ldots \phi^{i_1} \ldots \phi^{i_k} \ldots]
\eeq
is equivalent to $(J+\N)$ other terms in ${\tilde \Or}$
which are related to it by  cyclic permutations.
According to \cite{BMN}, at the leading order in the $1/J$ expansion, 
oscillators $y_n^i{}^\dagger$ correspond to 
insertions of $\phi^{i_k}$ with the phase $\exp({2 \pi i n_k a_k \over J})$,
where $a_k$ counts the number of $z$'s  to the left of this $\phi^{i_k}$.
One has to be be more careful when $1/J$ effects are taken into account.
In order for an operator to vanish when the level matching condition
is not satisfied, each sum over cyclically related terms in (\ref{co}) 
must vanish separately.
This happens when the phases assigned to the $\phi^{i_k}$ insertions are 
\beq
\label{qdef}
 q_{n_k}^{a_k} =  \exp\left( {2 \pi i n_k a_k \over J+\N}\right).
\eeq
Here $a_k$ counts {\it all} operators 
appearing to the left of the  $\phi^{i_k}$ insertion, 
and not just the $z$-s.
Similar arguments can be made to fix the form of $\Or_*$.
Again, each $\phi^{i_k}$ insertion comes with a phase
given by (\ref{qdef}).
In order to satisfy the level matching condition we should
also assign a phase $q_{n_k+n_l}^{a_{\bar z}}$ to $\bar z$.

To summarize, we have a correspondence which relates
SYM operators and plane wave string states away from the strict
Penrose limit
\beq
\label{opdef}
  \Or= ({\tilde \Or} -O_* ) 
\quad\lra\quad 
   y_{n_1}^{i_1}{}^\dagger \ldots y_{n_k}^{i_k}{}^\dagger \ldots |\eta \rangle,
\eeq
where
\beqa
 {\tilde \Or}
     &=& {1 \over \sqrt{\Omega} } \sum
   \left( \prod_k q_{n_k}^{a_k} \right) \tr[ z \ldots \phi^{i_1} \ldots \phi^{i_k} \ldots ], \\
 \Or_*&=& {1 \over \sqrt{\Omega} }
         \sum_{(n_p, n_q):  \,i_p=i_q}
      \left( \prod_{k \neq p,q} q_{n_k}^{a_k} \right) q_{n_p+n_q}^{a_{\bar z}} \,
         \tr[ z\ldots \phi^{i_1} {\ldots} \phi^{i_k} {\ldots}
                \check{\phi}^{i_k} {\ldots}  \check{\phi}^{i_l} 
                         {\ldots} {\bar z}\ldots],
\eeqa
and the phases $q_{n_k}^{a_k}$ are given by (\ref{qdef}).

The normalization constant $\Omega$ will be chosen so that
the leading term in $1/J$ expansion of the
$\Or(g^0)$ two-point function is normalized to one.
This leading term is given by the interaction-free diagrams
\beq
\label{intfree}
\III{\phi^{i_1}_{n_1}}{\phi^{i_1}_{n_1}}{}{}{\phi^{i_k}_{n_k}}{\phi^{i_k}_{n_k}},
\eeq
where the subscript $n_k$ in $\phi^{i_k}_{n_k}$ means that 
the corresponding insertion of $\phi^{i_k}$ in the string of
operators comes with the phase $q_{n_k}^{a_k}$.
Expression 
(\ref{intfree}) contains only contractions
of the same $\phi^{i_k}_{n_k}$.
Interaction-free diagrams with contractions of  $\phi^{i_k}_{n_k}$
and  $\phi^{i_l}_{n_l}$ with $n_k \neq n_l$ are also allowed, as
long as $i_k=i_l$.
Such diagrams however are subleading in $1/J$.

From (\ref{intfree}) 
we 
infer
that 
\beq
  \Omega = c N^{J+\N} (J+\N) \Ot,
\eeq
where $c$ is an irrelevant numerical prefactor; 
$N^{J+\N}$ arises
from the number of color loops in (\ref{intfree}); 
and $(J+\N)$ takes 
care of the fact that performing a cyclic permutation 
in one of the operators entering the two-point function gives an equivalent diagram.
When no oscillators are excited more than once, there is no further choice
of contractions and $\Ot$ is equal to the number of ways $\N$ 
$\phi$'s can be distributed among $J$ $z$'s, $\Ot=\prod_{n=1}^\N (J+n)$.
When there are multiple excitations of the same mode, 
there can be $N^i_n!$ inequivalent permutations of the 
$\phi^{i}_{n}$ in either one of the operators.
This gives rise to $N^i_n!$ copies of the diagram (\ref{intfree}).
We conclude that in general, 
\beq
  \Ot= \prod_k N^{i_k}_{n_k}! \prod_{n=1}^\N (J+n).
\eeq


\subsection{Two-point functions and the light cone Hamiltonian}
\label{section:anomalous dims}

The light cone energy of a string state and its momentum are related
to the anomalous dimension $\Delta$ and $R$-charge $J$ of the corresponding 
operator as follows
\beqa
\label{eq:H-lc-p's way}
  H_{lc}&=&
- \PP_+ ~~ = i {\partial \over \partial X^+} = 
\Delta-J, \\
\label{etaj}
  \eta&=& 
- \half \PP_- = {i\over2} {\partial \over \partial X^-} = 
{\Delta + J \over 2 R^2}.
\label{eq:eta-p's way}
\eeqa 
One can find anomalous dimensions of the gauge theory operators 
by looking at two-point functions, and we are now going to explain 
this in detail.
We will only consider planar diagrams. 
This amounts to neglecting 
string amplitudes of genus one and higher. 
Furthermore, we will only look at the terms in $H_{lc}$ which
behave like
\beqa
\label{pws}
 && {1 \over \eta^{2a}} \sim \left( {R^2 \over J} \right)^{2a} = {(4 \pi g N)^{a} \over J^{2a}},
                 \\ \nonumber
 && {1 \over R^2 \eta^{2 a+1}} \sim {1 \over R^2} \left( {R^2 \over J} \right)^{2a+1} =
                                              {(4 \pi g N)^a \over J^{2a+1}}, \\ \nonumber \\ \nonumber
           && \mbox{with $a = 0, 1$.}
\eeqa
On the string theory side, the first line in (\ref{pws}) corresponds to
the truncated expansion in powers of $1/\eta^2$ of the
plane wave Hamiltonian $H_0$.
The second line corresponds to the expansion of $H_1$.
Terms  in the two lines differ by a factor of $1/J$.
On the gauge theory side this factor arises when finite $J$ corrections
are taken into account, which leads to the modification of BMN operators,
explained in section \ref{section:operators}.
The first perturbative (from the SYM point of view) correction to
the light cone energy in (\ref{pws}) corresponds to $a=1$,
which implies that $a=0$ term in the second line of (\ref{pws}) vanishes.
This is in complete accord with the expansion of $H_1$ in powers of $1/\eta$.

Consider a set of gauge theory operators $\Or_\alpha$ 
labeled by $\alpha=\{(i_k,n_k)\}$. 
We will be interested in the SYM operators 
which correspond to plane wave states 
with $\N$ worldsheet oscillators excited.
Their two-point functions can be arranged as
\beqa
\label{2g}
  \langle \Or_\alpha(x) {\bar \Or}_\beta(0) \rangle &=&
 \langle \Or_\alpha(x) {\bar \Or}_\beta(0) \rangle_{g^0}
 +\langle \Or_\alpha(x) {\bar \Or}_\beta(0) \rangle_{g^1} + \OO(g^2) \\ \nonumber &=&
             {1 \over |x|^{2(J+\N)}} \left[
         {\bf T}_{\alpha\beta} - \F_{\alpha\beta} \log (\mu^2 x^2) 
+ \OO(g^2) 
\right].
\eeqa
Here, ${\bf T}$ is a matrix of combinatorial factors 
which come from interaction-free diagrams in  
$\langle \Or_\alpha(x) {\bar \Or}_\beta(0) \rangle_{g^0}$, 
while ${\bf F}$ captures the $\OO(g)$ effects of SYM interactions
in $\langle \Or_\alpha(x) {\bar \Or}_\beta(0) \rangle_{g^1}$. 
$\OO(g)$ contributions to the two point functions (\ref{2g}) 
come from diagrams of the type 
\beq
\label{4ptsample}
  \XII{}{}{}{}{}{}.
\eeq
In appendix \ref{section:feynman rules} we show that 
(\ref{4ptsample}) is equal to 
\beq
\label{eq:gamma value}
   \gamma \equiv -\beta \log \mu^2 x^2
          \equiv -{g N \over 2 \pi} \log \mu^2 x^2
\eeq
times a numerical factor 
determined by the fields which go into the 4-point vertex.

Operators $\OO_\alpha$ may not have well defined scaling dimensions 
at order $\OO(g)$. 
To find pure operators and their anomalous dimensions, 
we need to transform to a basis of eigenstates of the dilatation
operator.
By a linear transformation, 
we should bring (\ref{2g}) to the form 
\beq
   {1 \over |x|^{2(J+\N)}} \left[{\bf 1}-{\rm diag}[\{\lambda_\rho\}] 
	\log (\mu^2 x^2)\right],
\eeq
where the order $\OO(g)$ anomalous dimensions 
$\lambda_\rho$ are the eigenvalues of ${\bf T}^{-1} \F$, 
and ${\bf 1}$ is a unit matrix  
\cite{Ryzhov}. 
The matrices in (\ref{2g}) have the form
\beqa
\label{eq:T definition}
  {\bf T} &=& {\bf 1}+ {1\over J}{\bf T}^{(1)} 
+ \OO(1/J^2)
, \\
  \F      &=& \F^{(0)} +{1\over J} \F^{(1)}
+ \OO(1/J^2)
,
\label{ff}
\eeqa
since the operators $\OO_\alpha$ were chosen to be 
orthonormal at leading order, 
see the end of section \ref{section:operators}. 
Hence, up to corrections that are higher order in $1/J$
\beq
\label{g2form}
 { {\bf T}^{-1} \F }=  \F^{(0)}
     +{1 \over J} \left(  \F^{(1)} -  {\bf T}^{(1)}  \F^{(0)} \right)
+ \OO(1/J^2)
.
\eeq

Finally, light cone energies of worldsheet states 
are related to the 
quantum numbers of operators in \NN=4 SYM 
as 
\beq
\label{lcesym}
\Delta-J=\N+\lambda_\rho
.
\eeq
In other words, $\N\,{\bf 1}+{\bf T}^{-1} \F$, 
plays the role of the light cone Hamiltonian.
In the next section we will show that 
$H_{lc}-\N=H_0+H_1-\N$ 
is identical to ${\bf T}^{-1} \F$ computed in the
gauge theory 
[the $H_0$ and $H_1$ are given by
(\ref{h0e}), (\ref{h1diag}) and (\ref{h1offdiag})].
This means that to the accuracy we are working at, 
the spectrum of eigenstates of the light cone worldsheet Hamiltonian 
is the same as the spectrum of the dilatation operator 
in SYM.


\subsection{Equality of matrix elements} 
\label{section:equalitysimple}
Let us now show that ${\bf T}^{-1} \F$ and $H_{lc}-\N$ indeed
have the same matrix elements that are relevant for the
diagonalization.
In this section we 
consider matrix elements between states 
with all modes having distinct SO(4) indices.
We also assume that no modes are excited
more than once, $N^i_n \le 1$.
In appendix C we generalize these results to 
matrix elements between arbitrary plane wave states. 

The 
relevant off-diagonal terms in  $H_{lc}$ 
are given by (\ref{h1offdiag}).
When sandwiched between 
\beq
\label{i1}
  |\X\rangle  =  y_{n_1}^{i_1}{}^\dagger  \ldots  y_m^i{}^\dagger  y_n^j{}^\dagger  |\eta \rangle,
 \qquad   m \neq n,
\eeq
and
\beq
\label{f1}
  |\X'\rangle  =  y_{n_1}^{i_1}{}^\dagger  \ldots  y_m^j{}^\dagger  y_n^i{}^\dagger  |\eta \rangle,
  \qquad  m\neq n,
\eeq
with $i \neq j$, the off-diagonal part of the Hamiltonian 
(\ref{h1offdiag}) 
gives rise to the following matrix element
\beq
\label{del1}
  \langle \X| H_1^{OD}  |\X' \rangle  =  
      -{1 \over J} \left({ R^2 \over J}\right)^2 m n \sqrt{ N^i_m N^j_n N^i_n{}' N^j_m{}'},
\eeq
where we expressed $\eta$ as 
\beq
    \eta={J \over R^2} \left( 1 + {1 \over 2 J} \sum_{i,m} N^i_m 
+ \OO(1/J^2)
\right).
\eeq
This follows from (\ref{eq:H-lc-p's way}) and (\ref{eq:eta-p's way}). 
The second term in the brackets gives an $\Or(1/J)$
correction when used in the leading order Hamiltonian 
(\ref{h0e}). 
We should also reinstate the normal ordering term (\ref{not}).
The SYM calculations will fix it to be ${1 \over R^2 \eta^3} \sum_{i,n} n^2 N^i_n$.
Combining these contributions, the diagonal matrix elements%
\footnote{
	Here and below $\Or$ stands for an arbitrary
	worldsheet state or SYM operator, 
	for example $\X$ or $\X'$. 
	Diagonal matrix elements are all given 
	by the same expression.} 
read 
\beq
\label{hdiagx}
    \langle \Or| H_0{+}H_1^D{-}\N  |\Or \rangle  {=} 
        {1 \over 2} \left({R^2 \over J} \right)^2 \sum_{i,n} n^2 N^i_n
         {+}{1 \over J} \left({R^2 \over J} \right)^2 
\bigg[
{-}
\sum_{i,j,m,n} n^2 N^i_m N^j_n 
{+}         
\sum_{i,n} {n^2  N^i_n (N^i_n{+}1) \over 4} 
\bigg].
\eeq
%
For the states considered in this subsection $N^i_n=1$ and 
off-diagonal elements of $H_1$ other than (\ref{del1}) vanish.

We will also denote the SYM operators corresponding to 
states (\ref{i1})-(\ref{f1}) by $\X$ and $\X'$. 
As explained in section \ref{section:operators}, 
no terms of the type $\tr[ z \ldots {\bar z} \ldots \phi^i \ldots]$ 
appear as long as all $i_k$ labels distinct. 
That is, $\X_*=\X_*'=0$, and $\X={\tilde \X}, \X'={\tilde \X'}$.
Contributions to ${\bf T_{\X\X'}}$ and ${\bf T}_{\Or\Or}$
come from the diagrams  similar to (\ref{intfree}), 
\beq
\label{intfree2}
\III{\phi^i_m}{\phi^i_{m'}}{\phi^j_n}{\phi^j_{n'}}
 {\phi^{i_k}_{n_k}}{\phi^{i_k}_{n_k}}.
\eeq
\\
The top and bottom rows in (\ref{intfree2}) 
correspond to the two SYM operators entering the
two-point function.
Summing the phases over positions of $\phi$'s 
we obtain
\beq
\label{2ptcs}
    {1 \over \Ot} \sum{}'  \prod_k r_k^{a_k} = \delta_{mm'}\delta_{nn'}-   
                  { \delta_{m+n,m'+n'} (1- \delta_{mm'}\delta_{nn'})\over J}
+\OO(1/J^2). 
\eeq
The prime on the sum in (\ref{2ptcs}) means that we count 
modulo cyclic permutations, 
and we defined 
\beq
 r_{n_k} \equiv q_{n_k} q_{n_k'}{}^*=
     \exp\left({ 2\pi i ( n_k-n'_k) \over J+\N}\right).
\eeq
Only $r_m$ and $r_n$ are different from one, so 
(\ref{2ptcs}) can be computed by making use of
the invariance under cyclic permutations and fixing $a_m=0$ (and so $r_m^{a_m}=1$). 
The $\Or(1/J)$ term in (\ref{2ptcs}) appears because the range
of $a_n$ is $[1,J+\N-1]$.
Contributions with more than two $r_{n_k} \ne 1$ 
are suppressed by at least $1/J^2$ compared to (\ref{2ptcs}), 
so we do not need to worry about them.
Comparing (\ref{2ptcs}) with 
(\ref{2g}) and (\ref{eq:T definition}), 
we arrive at
\beq
\label{tods}
  {\bf T}^{(1)}_{\X\X}={\bf T}^{(1)}_{\X'\X'}=0, \qquad {\bf T}^{(1)}_{\X\X'}=-1.
\eeq

We now turn to the computation of $\F$.
Consider the diagrams that contribute both to 
$\langle \X(x) {\bar {\X'}}(0) \rangle_{g^1}$
and to the diagonal correlator $\langle \Or(x) {\bar \Or}(0) \rangle_{g^1}$. 
These are
\beq
\label{cr1}
 \IIX{}{}{}{}{\phi^{i_k}_{n_k}}{\bar z}{z}{\phi^{i_k}_{n_k}}+
\IIX{}{}{}{}{\phi^{i_k}_{n_k}}{\phi^{i_k}_{n_k}}{z}{\bar z} + (\phi^{i_k}_{n_k} \lra z, \phi^{i_k}_{n_k} \lra {\bar z}),
\eeq
and
\beq
\label{cr2}
\XIIm{\phi^{i_k}_{n_k}}{\phi^{i_k}_{n'_k}}{\phi^{i_l}_{n_l}}{\phi^{i_l}_{n'_l}}
           {z}{\bar z}{\phi^{i_k}_{n_k}}{\phi^{i_k}_{n_k}} +
\XIIm{\phi^{i_k}_{n_k}}{\phi^{i_l}_{n'_l}}{\phi^{i_l}_{n_l}}{\phi^{i_k}_{n'_k}}
               {z}{\bar z}{\phi^{i_k}_{n_k}}{\phi^{i_k}_{n_k}} + 
               (\phi^{i_k}_{n_k} \lra \phi^{i_l}_{n_l},\phi^{i_k}_{n_k'} \lra \phi^{i_l}_{n_l'} ).
\eeq
\\
The level matching condition gives 
\beq
\label{momc}
m+n=m'+n'.
\eeq
The diagrams in (\ref{cr2}) which contribute to 
${\bf F}^{(1)}_{\X \X'}$
have $n_k=n'_l=m, n_l=n'_k=n$.
The contribution (\ref{cr1}) differs from the interaction-free
diagram (\ref{intfree2}) just by an overall factor
\beq
\label{dfactor}
   -\gamma \, ( q_{n'_k}{}^*+q_{n'_k}-2 ).
\eeq
Therefore, 
summing over possible configurations of fields 
gives (\ref{2ptcs}) times (\ref{dfactor}), 
for a particular $\phi^{i_k}_{n_k}$ participating in the 
interaction vertex in (\ref{cr1}). 
Since any one of the $\phi^{i_k}_{n_k}$ can 
be used in the interaction (\ref{cr1}), 
this must be further summed over $k$. 
We find 
\beq
\label{f0c}
   -\gamma \left[\delta_{mm'}\delta_{nn'}-
   { \delta_{m+n,m'+n'} (1- \delta_{mm'}\delta_{nn'})\over J} \right]
        \sum_k \, ( q_{n'_k}{}^*+q_{n'_k}-2 ).
\eeq
This expression overcounts certain diagrams 
which do not appear in (\ref{cr1}). 
More precisely, whenever two $\phi$'s 
are sitting next to each other 
the $\phi-\phi$ line cannot cross or touch a $z-\bar z$ line, 
either to the left or to the right. 
We will deal with such diagrams separately.

We can read off the $\Or(J^0)$ part of $\F$ from (\ref{f0c}) 
by using (\ref{2g}) and $\gamma=-\beta \log \mu^2 x^2$, 
\beq
\label{f0u}
   \F^{(0)}=-\beta \sum_k \, ( q_{n_k}{}^*+q_{n_k}-2 ).
\eeq
Expanding the $q$'s  in powers of  $1/J$ and taking the
leading term gives the result of BMN,
\beqa
\label{f0lo}
  \F^{(0)}&=& f^{(0)} {\bf 1}, \\ 
  f^{(0)}&=&{2 \pi g N \over J^2} \sum_k (n_k)^2 = 
             {1 \over 2} \left( R^2 \over J \right)^2 \sum_n n^2 N^i_n.
\eeqa
To get $\Or(1/J)$ corrections to this result we have to
be more careful.
As explained above, in (\ref{f0c}) we overcounted 
the configuration of fields where 
two $\phi$'s appear next to one another 
in the top row of (\ref{cr1}), as in (\ref{cr2}): 
\beq
\label{nal}
(q_{n_k}^{a_k} q_{n_l}^{a_l} \ldots)  \tr[ \ldots \phi^{i_k} \phi^{j_l} \ldots]+(k\lra l).
\eeq
Now diagrams in (\ref{cr1}) where $\phi^{i_k}$ ($\phi^{j_l}$) interacts with 
$z-\bar z$ propagator to the right (left) are not allowed.
The value of such diagrams is
\beq
\label{vdo}
  -\gamma ( q_{n'_k}^*+q_{n'_l}-2) q_{n_l} q_{n_l'}^* \,
     r_{n_k{+}n_l{-}n'_k{-}n'_l}^{a_k}  +(k\lra l)=
       -\gamma (q_{n_k}^*+q_{n_l}-2  q_{n_l} q_{n_l'}^*) \,
     r_{n_k{+}n_l{-}n'_k{-}n'_l}^{a_k}  +(k\lra l),
\eeq
where we used (\ref{momc}).
Their contributions have to be substituted by the ones that appear
in (\ref{cr2}) instead. 
These are given by
\beq
\label{vdc}
   \gamma ( q_{n_l} q_{n_l'}^* - q_{n_l} q_{n'_k}^* ) r_{n_k{+}n_l{-}n'_k{-}n'_l}^{a_k}+(k\lra l).
\eeq
The difference of (\ref{vdc}) and (\ref{vdo}) is the same
for both diagonal ($n'_k=n_k, n'_l=n_l$) and off-diagonal 
($n'_k=n_l, n'_l=n_k$) cases, 
and equals
\beq
\label{add1}
   \gamma (q_{n_k}^*+q_{n_l}- q_{n_l} q_{n_k}^* -1 )  r_{n_k{+}n_l{-}n'_k{-}n'_l}^{a_k}
       +(k\lra l).
\eeq
This should be summed over $a_k$ and divided by the normalization constant $\Ot$.
Since the number of configurations with two $\phi$'s next to each other is 
down by $1/J$ compared to the total number of configurations, 
we pick up an overall factor of $1/J$. 
Configuration which have three and more $\phi$'s
next to each other are suppressed by even higher powers of $1/J$, 
and we can neglect them to the order we are working. 

The full result for 
$\langle \X(x) {\bar \X'}(0)\rangle_{g^1}$ is given by (\ref{f0c}), 
plus (\ref{add1}) with $n_k=m, n_l=n$.
Other terms in (\ref{cr2}) are $\Or(1/J^2)$ and are not important for us.
Since $m'=n, n'=m$ for an off-diagonal element, the
first term in (\ref{f0c}) vanishes, and we have
\beq
\label{fxod}
 \F_{\X\X'}^{(1)}
= -\beta   \left[ \sum_k \, ( q_{n'_k}{}^*{+}q_{n'_k}{-}2 )+
              (q_{m-n}^*{+}q_{m-n}{-}q_m^* {-} q_m {-}q_n^*{-}q_n {+}2 ) \right].
\eeq
To get the corresponding off-diagonal element of the light cone Hamiltonian, 
we should add 
$-[{\bf T}^{(1)} \F^{(0)}]_{\X\X'}/J$
to $\F_{\X\X'}^{(1)}/J$, 
see (\ref{g2form}). 
According to (\ref{tods}) and (\ref{f0u}), such addition precisely
cancels the first term in (\ref{fxod}), 
and we find 
\beq
  [{\bf T}^{-1} \F]_{\X\X'}= -{\beta   \over J}
                  (q_{m-n}^*{+}q_{m-n}{-}q_m^* {-} q_m {-}q_n^*{-}q_n {+}2 ).
\eeq       
Expanding the $q$'s in powers of $1/J$, taking the leading term and
substituting the value of $\beta$ we arrive at
\beq
\label{del1a}
   [{\bf T}^{-1} \F]_{\X\X'}= -{1 \over J} \left({R^2 \over J}\right)^2 m n.
\eeq
This reproduces the string theory off-diagonal matrix element
(\ref{del1}), since for the states we are considering $N^i_n=1$.

Let us now compute the diagonal terms.
Now all of the diagrams in (\ref{cr2}) contribute, (\ref{add1})
should be summed over $k$ and added to (\ref{f0c}) with
$m'=m, n'=n$.
This gives
\beq
\label{diagmee}
 \langle \Or(x)\Or(0)\rangle_{g^1}=  -\gamma \sum_k \, ( q_{n_k}{}^*+q_{n_k}-2 )
                          +{\gamma \over 2 J} \sum_{k\neq l}
                          (q_{n_k}{+}q_{n_l}- q_{n_l} q_{n_k}^* -1 +c.c.).
\eeq       
Since ${\bf T}_{\Or\Or}=1$, we have
\beq
\label{diagme}
   [{\bf T}^{-1} \F]_{\Or\Or}=  -\beta \sum_k \, ( q_{n_k}{}^*+q_{n_k}-2 )
                          +{\beta \over 2 J} \sum_{k\neq l}
                          (q_{n_k}{+}q_{n_l}- q_{n_l-n_k} -1 +c.c.).
\eeq
The first term gives (\ref{f0lo}) at the leading order, 
however the definition (\ref{qdef}) of $q_{n_k}$ implies
that there is a $1/J$ correction to the leading term.
Expanding in powers of $1/J$ and keeping terms up to $\Or(1/J)$
one can write (\ref{diagme}) as
\beq
\label{tfdiag}
   [{\bf T}^{-1} \F]_{\Or\Or}={1 \over 2} \left({R^2 \over J} \right)^2 \sum_n n^2  +
         {1 \over J} \left({R^2 \over J} \right)^2 \left(
         {-}\sum_{i,j,m,n} n^2 N^i_m N^j_n 
         {-}{1 \over 2} \sum_{k \neq l:\, i_k{\neq} i_l} n_k n_l  \right).
\eeq
Using the level matching condition 
(which now reads $\sum_k n_k=0$), 
we can write the last term in parenthesis as
\beq
  {-}{1 \over 2} \sum_{k \neq l:\, i_k{\neq} i_l} n_k n_l  =
     {1 \over 2} \sum_{k} n_k^2.
\eeq
Substituting this back into (\ref{tfdiag}) one can see that the
resulting expression is equal to the string theory result (\ref{hdiagx}).
In appendix \ref{section:equalitygeneric} we generalize
the results of this subsection to matrix elements between the
generic states.


\section{Summary and further developments}
\label{section:discussion}

It has been known for some time \cite{Metsaev,Metsaev:2002re} 
that type IIB string theory is solvable in the plane wave background,
which can be viewed as the Penrose limit of $AdS_5 \times S^5$.
BMN \cite{BMN} showed that the string spectrum in this background,
can be recovered from the boundary \NN=4 super Yang-Mills.
Motivated by these results, we analyzed the properties of
this correspondence when finite radius effects are included.
We found that to the leading order in $1/R^2$ and $\lambda'=g N/J^2$, 
the string theory spectrum matches the spectrum
of anomalous dimensions of (linear combinations of) BMN operators.
%
On the string side we have an interacting worldsheet 
theory, when the leading $\Or(1/R^2)$ corrections
to the plane wave metric are taken into account.
Leading corrections to the string spectrum can then be computed
with quantum mechanical perturbation theory.
On the SYM side, departing from the Penrose limit forces one to refine
the BMN operators, paying attention to $1/J$ corrections.
We nevertheless assume that these refined operators continue
to correspond to plane wave states even away from the Penrose limit.
Such operators however do not have definite scaling dimensions,
when $1/J$ corrections are included.
Finding the spectrum of scaling dimensions in SYM requires
one to compute the matrices of two-point functions
$\langle \Or_\alpha {\bar \Or}_\beta \rangle_{g^0} \sim {\bf T}_{\alpha\beta}$
and 
$\langle \Or_\alpha {\bar \Or}_\beta \rangle_{g^1} \sim \F_{\alpha\beta}$.
Then, ${\bf T}^{-1}\F$ 
is related to the light cone worldsheet Hamiltonian. 
We find matching between the matrix elements of these operators.

There is a number of questions raised by the results of this paper.
It would be interesting to see if the correspondence between 
the operators we define in section \ref{section:operators} and
plane wave states is exact and holds for arbitrary values of AdS radius.
So far we matched the leading $1/R^2$, $\lambda'$ terms in
matrix elements of the light cone Hamiltonian.
We did not include the fermionic part of the superstring
in our analysis, which led to an undefined normal ordering
constant in diagonal matrix elements.
Incorporating fermions and extending the results of 
\cite{Metsaev,Metsaev:2002re}
to $\Or(1/R^2)$ corrections is an interesting open problem.
It would also be interesting to extend our analysis 
to higher powers of $\lambda'$.
This would require computing diagrams with multiple interactions,
but perhaps one may be able to come up with a resummation
technique similar to the one introduced in \cite{BMN}.
Extending our results to higher orders in $1/R^2$ seems 
more difficult technically, but might also deserve some interest.

Other possible extensions include studying backgrounds
that are more complicated than $AdS_5 \times S^5$.
Probing the strong coupling behavior of boundary
theories with fewer supersymmetries may be of
particular interest, but it remains to be seen how far one can go 
with this perturbative approach.


\section{Acknowledgments}

We would like to thank Per Kraus, David Kutasov and David Sahakyan
for many useful discussions, as well as for carefully reading 
the preliminary version of the manuscript. 
We are grateful to Dan Freedman and Lubos Motl 
for sharing their insights on their construction of 
leading order SYM operators. 
The work of A.P. was supported by DOE grant \#DE-FG02-90ER40560.


\pagebreak

{\LARGE \bf \appendixname}

\appendix


\section{An alternative worldsheet discussion}
\label{section:gsw way}

In Section \ref{section:worldsheet} we discussed 
how to do the worldsheet calculations in the 
spirit of Polchinski \cite{Polch}. 
In this Appendix, we explain in detail 
how to fix the gauges using the method described 
in GSW \cite{GSW}. 
We find the same results for physical quantities 
as in Section \ref{section:worldsheet}.

\subsection{Penrose limit of $AdS_5 \times S^5$}
\label{section:leading order}

Before fixing any gauges, 
the bosonic part of the worldsheet action is 
\begin{eqnarray}
\label{eq:basic action:again}
S = - {1\over 4 \pi \alpha'} 
\int (d^2 \sigma) \sqrt{-\gamma} \gamma^{ab} G_{ab}
\end{eqnarray}
where 
the induced metric on the worldsheet is 
$G_{ab} \equiv \partial_a X^\mu \partial_b X^\nu G_{\mu\nu}$. 
Using reparametrization invariance and 
Weyl invariance, we can bring the worldsheet metric to the 
form 
\begin{eqnarray}
\label{eq:metric:gamma=eta}
\gamma_{ab} = \eta_{ab} = 
\left(\matrix{
-1 & 0 \cr 0 & 1
}\right)
\end{eqnarray}
in $(\tau,\sigma)$ coordinates. 
The leading order target space metric (\ref{eq:metric:leading order}) is 
\begin{eqnarray}
\label{eq:metric:leading order:again}
ds^2_0 = - 4 dX^- dX^+ - (r^2+y^2) dX^+ dX^+ + dX_I \, dX_I 
\end{eqnarray}
$I=1, ... , 8$. 
After fixing the worldsheet metric as in (\ref{eq:metric:gamma=eta}), 
the string action (\ref{eq:basic action:again}) becomes 
\begin{eqnarray}
\label{eq:action:leading}
S_0 &=& - {1\over 4 \pi \alpha'} 
\int (d\tau d\sigma) 
\left[
4 \dot X^- \dot X^+ + X^2 \dot X^+ \dot X^+ - \dot X_I \dot X_I 
\right.\nonumber\\&&\hspace{6em}\left.
- 4 (X^-)' (X^+)' - X^2 (X^+)' (X^+)' + X_I' X_I' 
\right]
\end{eqnarray}
%
The action (\ref{eq:action:leading}) is not completely gauge fixed. 
We still have the freedom to reparameterize 
the worldsheet coordinates holomorphically, 
\begin{eqnarray}
\label{eq:holomorphic:general}
\sigma^+ \to \tilde\sigma^+(\sigma^+)
,\quad
\sigma^- \to \tilde\sigma^-(\sigma^-)
\end{eqnarray}
where $\sigma^\pm = \tau \pm \sigma$ 
are the holomorphic and antiholomorphic 
worldsheet coordinates. 
Under 
(\ref{eq:holomorphic:general}), 
the new 
\begin{eqnarray}
\label{eq:holomorphic:new tau}
\tilde\tau = 
{1\over2} 
\left[
\tilde\sigma^+(\tau+\sigma) + \tilde\sigma^-(\tau-\sigma)
\right]
\end{eqnarray}
satisfies the free massless wave equation 
\begin{eqnarray}
\label{eq:holomorphic:new tau:wave eq}
\ddot{\tilde\tau} - \tilde\tau'' 
\equiv 
\left[
\partial_\tau^2 - \partial_\sigma^2 
\right] 
\tilde\tau 
= 0
\end{eqnarray}
$X^-$ enters the action (\ref{eq:action:leading}) linearly, 
so we can integrate it out, imposing its equation of motion 
as a constraint. 
This equation is 
$\ddot X^+ - (X^+)'' = 0$, 
and it has the form (\ref{eq:holomorphic:new tau:wave eq}). 
Hence we can choose the 
light-cone gauge 
\begin{eqnarray}
\label{eq:light-cone:X+}
X^+ = x^+ + p^+ \tau 
\end{eqnarray}
This exhausts all the gauge freedom in the problem. 
After integrating out $X^-$ and choosing the lightcone gauge, 
the action becomes 
\begin{eqnarray}
\label{eq:action:leading:lc}
S_0 &=& - {1\over 4 \pi \alpha'} 
\int (d\tau d\sigma) 
\left[
(p^+)^2 X^2 - \dot X_I \dot X_I + X_I' X_I' 
\right]
\end{eqnarray}
From this, 
we find the lightcone Hamiltonian to be 
\begin{eqnarray}
\label{eq:hamiltonian:leading}
H_0 
&=& {1\over 4 \pi \alpha'} \int_0^{2\pi} d\sigma 
\left[
(2 \pi \alpha')^2 P_I P_I + X_I' X_I' + (p^+)^2 X_I X_I 
\right]
\end{eqnarray}
where $P_I$ are the momenta conjugate to $X_I$. 
%
The Hamiltonian (\ref{eq:hamiltonian:leading}) is quadratic, 
and can be quantized exactly. 
Expand the $X_I$ and $P_I$ in modes as 
\begin{eqnarray}
\label{eq:oscillators}
X_I = 
\sum_{n=-\infty}^{+\infty} 
i \sqrt{\alpha' \over 2\varpi_n} 
\left[
a_n^I - a_n^I{}^\dagger
\right]
,\quad
2 \pi 
P_I = 
\sum_{n=-\infty}^{+\infty} 
\sqrt{\varpi_n \over 2 \alpha'} 
\left[
a_n^I - a_n^I{}^\dagger
\right]
\end{eqnarray}
where the frequencies are 
\begin{eqnarray}
\label{eq:frequencies}
\varpi_n = \sqrt{(p^+)^2 + n^2}
\end{eqnarray}
and the oscillators 
\begin{eqnarray}
\label{eq:oscillators:def}
a_n^I = \alpha_n^I e^{- i ( \varpi_n \tau - n \sigma )} 
,\quad
a_n^I{}^\dagger = 
\alpha_n^I{}^\dagger e^{+ i ( \varpi_n \tau - n \sigma )} 
\end{eqnarray}
close as 
$[ \alpha_m^I , \alpha_n^J{}^\dagger ] = \delta^{IJ} \delta_{mn}$. 
%
In terms of these oscillators, 
(\ref{eq:hamiltonian:leading}) reads 
\begin{eqnarray}
\label{eq:hamiltonian:leading:oscillators}
H_0 &=& 
\sum_{I=1}^8 \sum_{n=-\infty}^{+\infty}
\varpi_n 
\left[
N_n^I + {1\over2} 
\right] 
\end{eqnarray}
where the number operators are 
$N_n^I \equiv a_n^I{}^\dagger a_n^I$ 
(no sum on either $n$ or $I$). 
We will drop the normal ordering constants, 
since they cancel against the fermionic ones in the plane wave limit.


To compare space-time quantum numbers with worldsheet quantities, 
we look at the Noether charges associated with target space isometries. 
The relevant ones for us will be the energy 
$E = i \partial_t$, and the angular momentum $J = - i \partial_\psi$, 
where $t$ and $\psi$ are the global coordinates on $AdS$ 
used in (\ref{eq:exact metric}). 
In the dual CFT description, these correspond to the conformal dimension 
$\Delta = E$ and the $R$-charge $J$. 
We find 
\begin{eqnarray}
\label{eq:charges:general}
- i 
{\partial \over \partial X^\pm} 
\quad \lra \quad
\PP_\pm = 
\int_0^{2\pi} d\sigma P_\pm 
\label{eq:momenta:general}
,\quad\mbox{where } 
P_\mu = 
{
\delta S
\over 
\delta \dot X^\mu
}
= {1\over 2 \pi \alpha'} \, G_{\mu\nu}\dot X^\nu
\end{eqnarray}
are the momenta canonically conjugate to the coordinates $X^\mu$. 
In the light-cone gauge, 
\begin{eqnarray}
\label{eq:charges:leading:lc:start}
\label{eq:charges-st:leading:lc:start}
{\Delta + J \over R^2} 
\quad \lra \quad 
i 
{
\partial
\over 
\partial X^-
}
~\lra~
- \PP_- &=& 
{2 p^+ \over \alpha' }
\\
\Delta - J 
\quad \lra \quad 
i 
{
\partial
\over 
\partial X^+
}
~\lra~
- \PP_+ &=& 
{1\over p^+} H_0 
= 
\sum_I \sum_n 
{\varpi_n \over p^+} 
N_n^I 
\label{eq:charges-st:leading:lc:end}
\label{eq:H=P+}
\label{eq:charges:leading:lc:end}
\end{eqnarray}
Given our gauge choice (\ref{eq:light-cone:X+}), 
$\PP_+$ and $H_0$ should differ by a factor of $-p^+$; 
the minus sign in (\ref{eq:H=P+}) comes about because 
$H = i \partial_t$,  
while 
$P = - i \partial_X$. 
The light-cone states 
\begin{eqnarray}
\label{eq:states:def:w/s:first}
| I_m , J_n , ... \rangle 
\equiv
a^I_m{}^\dagger a^J_n{}^\dagger \; ... \; 
|0, p^+ \rangle
\end{eqnarray}
have 
$p^+ = {J \over \sqrt{4 \pi g N}}$, 
with $R^4 = 4 \pi g N \alpha'^2$; 
$\Delta - J = 
[1 + {m^2 \over 2(p^+)^2}] + [1 + {n^2 \over 2(p^+)^2}] + ... $ 
for large $p^+$.

Oscillators 
(\ref{eq:oscillators:def}) 
explicitly depend on time, so they are Heisenberg picture
operators. To go to the Schroedinger picture, we can just 
drop the time dependence and use the 
equations of motion which follow from 
the Hamiltonian
(\ref{eq:hamiltonian:leading}). 
These are 
\begin{eqnarray}
\label{eq:oscillators:schroedinger:def:start}
a_n^I &=& \alpha_n^I ~ e^{+ i n \sigma } 
,\quad
{d\over dt} a_n^I = - i \varpi_n \alpha_n^I ~ e^{+ i n \sigma } 
\\
\label{eq:oscillators:schroedinger:def:end}
a_n^I{}^\dagger &=& \alpha_n^I{}^\dagger e^{ - i n \sigma } 
,\quad
{d\over dt} a_n^I{}^\dagger 
= + i \varpi_n \alpha_n^I{}^\dagger ~ e^{- i n \sigma } 
\end{eqnarray}
It will be convenient to work with Heisenberg 
picture operators 
throughout, and convert the final expressions to 
the Schroedinger picture 
before doing 
perturbation theory.

\subsection{
Corrections to the Penrose limit of $AdS_5 \times S^5$} 
\label{section:corrections}

The $1/R^2$ correction to the space-time metric is 
given by ${1\over R^2} ds_1^2$ with 
\begin{eqnarray}
\label{eq:metric:correction:again}
ds_1^2 = 
- 2 dX^- dX^+ (r^2-y^2) - {1\over3} (r^4-y^4) dX^+ dX^+ + 
{1\over3} (r^4 d\Omega_3{}^2 - y^4 d\Omega'_3{}^2) 
\end{eqnarray}
%
Using the identities
$d r_i d r_i = dr^2 + r^2 d\Omega_3^2$ and 
$r d r = r_i d r_i$, 
we can write 
$r^4  d\Omega_3^2 = 
\left[
r_i r_i dr_j dr_j - r_i r_j dr_i dr_j
\right]$
and similarly for the $y$-s. 
This results in the contributions 
\begin{eqnarray}
\label{eq:X-ab}
X_{ab} \equiv {1\over 3} 
\left[
X_i X_i \; (\partial_a X_j) (\partial_b X_j)
-
X_i X_j \; (\partial_a X_i) (\partial_b X_j)
\right]
\end{eqnarray}
to the induced metric $G_{ab}$. 
$X$ can be either $r$ or $y$ in (\ref{eq:X-ab}), 
and the sums on the repeated $i$ and $j$ run from 1 to 4. 
The $i=j$ terms cancel in (\ref{eq:X-ab}).

After fixing the worldsheet metric as in (\ref{eq:metric:gamma=eta}), 
the bosonic part of the 
action 
becomes 
$S = S_0 + {1\over R^2} S_1$ with 
$S_0$ given in (\ref{eq:action:leading}), and 
\begin{eqnarray}
\label{eq:action:correction}
S_1 &=& - {1\over 4 \pi \alpha'} 
\int (d\tau d\sigma) 
\left\{
- 2 X^-
\left[ 
\partial_\tau (\dot X^+ ( r^2 - y^2 ) ) 
- 
\partial_\sigma ((X^+)' ( r^2 - y^2 ) ) 
\right] 
\right.\nonumber\\&&\hspace{6.5em}\left.
+ 
{1 \over 3} 
\left[ 
\dot X^+ \dot X^+ - (X^+)' (X^+)' 
\right]
(r^4 - y^4) 
- \left( r_{\tau\tau} - y_{\tau\tau} \right) 
+ \left( r_{\sigma\sigma} - y_{\sigma\sigma} \right) 
\right\}
\nonumber\\
\end{eqnarray}
(we integrated by parts so that 
derivatives of $X^-$ do not appear in $S_1$). 
Since the variable $X^-$ appears linearly in the action $S$, 
we can integrate it out, and impose its equation of motion 
as a constraint. 
Although this equation is no longer linear, 
it can be solved perturbatively in $1/R^2$. 
Writing 
\begin{eqnarray}
\label{eq:X+:breakdown}
X^+(\tau,\sigma) = 
X^+_0 
+ {1\over R^2} 
X^+_1 
\end{eqnarray}
where $X^+_{0,1}$ are both of order one, we get 
\begin{eqnarray}
\label{eq:eom:correction:linearized}
0 = 
\ddot{X}^+_0 
- 
(X^+_0)'' 
+ {1\over R^2} 
\left\{
\ddot{X}^+_1 
- (X^+_1)'' 
+ \partial_\tau 
\left[
\dot X^+_0 
\left( {r^2 - y^2 \over 2} \right) 
\right] 
- \partial_\sigma 
\left[
(X^+_0)' 
\left( {r^2 - y^2 \over 2} \right) 
\right] 
\right\} 
\hspace{-1em}\nonumber\\
\end{eqnarray}
Since $X^+_0$ satisfies the 
free massless wave equation, we can take 
$X^+_0 = x^+ + p^+ \tau$. 
Thus the (modified) light-cone gauge choice is 
\begin{eqnarray}
\label{eq:lc gauge}
X^+(\tau,\sigma) = 
\left(
x^+ + p^+ \tau 
\right)
+ {1\over R^2} 
X^+_1 
\end{eqnarray}
To completely fix the gauge, we have to make sure that 
contributions of the form 
$x_1^+ + p_1^+ \tau$ and $e^{i n (\tau \pm \sigma)}$, 
are absent in the mode expansion of $X_1^+(\tau,\sigma)$. 
In terms of the original coordinate $X^+$ and the 
original $\tau$ and $\sigma$, this is a statement 
that 
\begin{eqnarray}
\label{eq:X1+:alternative def}
{1\over R^2} 
X_1^+(\tau,\sigma) \equiv X^+(\tau,\sigma) 
- {1\over 2\pi} \sum_{n, \pm} e^{i n (\tau \pm \sigma)} 
\int (d\sigma d\tau) X^+(\tau,\sigma) e^{- i n (\tau \pm \sigma)}
\end{eqnarray}

The leftover piece $X_1^+$ is not a new dynamical variable; 
rather, it depends on $r_i$ and $y_i$. 
It is defined to satisfy 
\begin{eqnarray}
\label{eq:eom:linearized}
\ddot{X}^+_1 
- (X^+_1)'' 
+ {1 \over 2} p^+ 
\partial_\tau 
\left( r^2 - y^2 \right) 
= 0
\end{eqnarray}
The $r^2$ and $y^2$ should be taken as their leading 
order versions (\ref{eq:oscillators}). 
Setting 
$r^i_n \equiv a^i_n$ and $y^i_n \equiv a^{i+4}_n$ 
in the mode expansions (\ref{eq:oscillators}), 
we find 
\begin{eqnarray}
\label{eq:X+ correction}
X^+_1 &=& 
{i p^+ \alpha' \over 2} 
\sum_{m \, , \; n} 
{ \varpi_n \over \sqrt{\varpi_m \varpi_n } }
{ 
\left[
( r_m^i r_n^i - r_m^i{}^\dagger r_n^i{}^\dagger ) - 
( y_m^i y_n^i - y_m^i{}^\dagger y_n^i{}^\dagger )
\right]
\over 
\left[(\varpi_m + \varpi_n)^2 - (m+n)^2 \right]
} 
\nonumber\\&+& 
{i p^+ \alpha' \over 2} 
\sum_{m \ne n} 
{ \varpi_n \over \sqrt{\varpi_m \varpi_n } }
{ 
\left[
( r_m^i r_n^i{}^\dagger - r_m^i{}^\dagger r_n^i ) - 
( y_m^i y_n^i{}^\dagger - y_m^i{}^\dagger y_n^i ) 
\right]
\over 
\left[(\varpi_m - \varpi_n)^2 - (m-n)^2 \right] 
} 
\end{eqnarray}
Equation (\ref{eq:eom:linearized}) 
is solved in Heisenberg picture; 
the operator $X^+_1$ is determined in terms of 
the Heisenberg picture oscillators 
(\ref{eq:oscillators:def}). 
Since (\ref{eq:X+ correction}) 
contains no explicit time dependence, 
it can be interpreted as a Schroedinger picture expression 
(when the oscillators are taken to be in Schroedinger picture), 
and used in perturbative calculations of energies. 

The action in the modified lightcone gauge (\ref{eq:lc gauge}) reads 
\begin{eqnarray}
\label{eq:action:correction:lc:again}
S &=& 
- {1\over 4 \pi \alpha'} 
\int (d\tau d\sigma) 
\left\{
(p^+)^2 (r^2 + y^2) 
- \dot r_i \dot r_i 
- \dot y_i \dot y_i 
+ r_i' r_i' 
+ y_i' y_i' 
\right.\hspace{-3em}\nonumber\\&&\hspace{3em}\left.
+ {1\over R^2} 
\left[ 
\left( r_{\sigma\sigma} - y_{\sigma\sigma} \right) 
- \left( r_{\tau\tau} - y_{\tau\tau} \right) 
+ {1\over 3} (p^+)^2 ( r^4 - y^4 ) 
+ 2 p^+ \dot X_1^+ (r^2 + y^2) 
\right] 
\right\}
\nonumber\\
\end{eqnarray}
after integrating out $X^-$, i.e. 
after solving the constraint equation (\ref{eq:eom:correction:linearized}). 
As discussed in \cite{p1}, 
the first order correction to the Hamiltonian 
is minus the correction to the Lagrangian, 
$\delta H = - \delta L$. 
%
Hence 
the (modified-)lightcone Hamiltonian is 
\begin{eqnarray}
\label{eq:hamiltonian}
H &=& 
{1\over 4 \pi \alpha'} \int_0^{2\pi} \!\! d\sigma 
\left\{
\left[
(2 \pi \alpha' )^2
( P^r_i P^r_i + P^y_i P^y_i )
+
( r_i' r_i' + y_i' y_i' )
+ (p^+)^2 
( r^2 + y^2 )
\right]
\right.\hspace{-2em}\nonumber\\&&\left.\hspace{1.4em}
+ 
{1\over R^2} 
\left[
( r_{\sigma\sigma} - y_{\sigma\sigma} ) 
- ( r_{\tau\tau} - y_{\tau\tau} ) 
+ {1\over 3} (p^+)^2 
( r^4 - y^4 ) 
+ 2 p^+ \dot X^+_1 ( r^2 + y^2 ) 
\right]
\right\}
\hspace{3em}
\end{eqnarray}
with $X_1^+$ given in (\ref{eq:X+ correction}). 
%

The conserved charges corresponding to 
${\Delta + J \over R^2}$ 
and 
$\Delta - J$
are 
\begin{eqnarray}
\label{eq:charges-st:correction:lc:start}
\label{eq:charges:correction:start}
-\PP_- &=& 
{2 p^+\over \alpha'} 
+ {4 p^+ \over 4 \pi \alpha'} \int_0^{2\pi} d\sigma \; 
{1\over R^2} \left( {\dot X_1^+ \over p^+} + {r^2 - y^2 \over 2} \right) 
\\
-\PP_+ &=& 
{1\over p^+} H
\label{eq:charges:correction:end}
\label{eq:charges-st:correction:lc:end}
\end{eqnarray}
%
In terms of the (Schroedinger picture) oscillators, 
\begin{eqnarray}
\label{eq:charges-st:correction:computed:start}
- \PP_- &=& 
{2 p^+\over \alpha'} 
\left\{
1 + {\alpha' \over 2 R^2} 
\left[ 
\sum_{i=1}^4 \sum_{n=-\infty}^{+\infty}
{1\over \varpi_n} (N_n^r{}^i - N_n^y{}^i)
\right] 
\right\}
\end{eqnarray}
Corrections of the form $a a$ and $a^\dagger a^\dagger$ 
precisely cancels 
between ${1\over p^+} \dot X_1^+$ and ${1\over 2}(r^2 - y^2)$ 
in (\ref{eq:charges-st:correction:lc:start}). 
For $p^+ \gg 1$, 
the worldsheet parameter $p^+$ is related to $J$ and $N$ as 
\begin{eqnarray}
\label{eq:ws p+ in terms of st J}
p^+ &=& 
{J \over\sqrt{ 4 \pi g N}} 
\left\{
1 + {1 \over J} 
\sum_{i=1}^4 \sum_{n=-\infty}^{+\infty}
\left[ N_n^y{}^i + {2 \pi g N n^2 \over J^2} N_n^r{}^i \right]
\right\}
\end{eqnarray}
to order $1/R^2$. 
Here we used $R^4 = 4 \pi g N \alpha'^2$, 
and wrote $(\Delta+J) = 2 J + (\Delta-J)$. 
In (\ref{eq:ws p+ in terms of st J})
the contributions of the $y$ and $r$ oscillators 
have rather different structure.

The Hamiltonian (\ref{eq:hamiltonian}) is relatively involved, 
so we analyze it in more detail. 
The leading order lightcone string states 
\begin{eqnarray}
\label{eq:states:def:w/s}
| a_p , b_q , ... \rangle 
= 
y^a_p{}^\dagger y^b_q{}^\dagger \; ... \; 
|0, p^+ \rangle
\end{eqnarray}
with worldsheet momenta 
$(p, q , ... )$ and $(p', q', ... )$ 
are degenerate only when the 
$(p, q, ... )$ and $(p', q', ... )$ 
are permutations of one another. 
Hence the only terms in $\delta H$ relevant 
for computing the first correction to 
the worldsheet energies, 
are the ones which permute the worldsheet momenta, 
namely 
$a_k a^\dagger_k a_k a^\dagger_k$ 
and 
$a_k a^\dagger_k a_l a^\dagger_l$.

Such terms in 
$\left[ (r_{\sigma\sigma}-y_{\sigma\sigma})
-(r_{\tau\tau}-y_{\tau\tau}) 
+ {1\over 3} (p^+)^2 (r^4-y^4) \right]$ 
combine as 
\begin{eqnarray}
\label{eq:terms from X-ss X-tt X4}
&&
2 \left({\alpha' \over 2} \right)^2 
\sum_k {(p^+)^2 \over \varpi_k^2} 
[ r_k^i r_k^i r_k^j{}^\dagger r_k^j{}^\dagger - 
y_k^i y_k^i y_k^j{}^\dagger y_k^j{}^\dagger]
\nonumber\\&+&
2 \left({\alpha' \over 2} \right)^2 
\sum_{k \ne l} {(p^+)^2 + \varpi_k \varpi_l - k l \over \varpi_k \varpi_l} 
[ r_k^i r_l^i r_l^j{}^\dagger r_k^j{}^\dagger - 
y_k^i y_l^i y_k^l{}^\dagger y_k^j{}^\dagger]
\nonumber\\&+&
2 \left({\alpha' \over 2} \right)^2 
\sum_{k \ne l} {(p^+)^2 - \varpi_k \varpi_l + k l \over \varpi_k \varpi_l} 
[ r_k^i r_l^j r_l^i{}^\dagger r_k^j{}^\dagger - 
y_k^i y_l^j y_l^i{}^\dagger y_k^j{}^\dagger]
\end{eqnarray}
and the term $2 p^+ \dot X_1^+ (r^2 + y^2)$ gives 
\begin{eqnarray}
\label{eq:terms from X-dot X-dot}
&-&
2 \left({\alpha' \over 2} \right)^2 
\sum_k 
[ r_k^i r_k^i r_k^j{}^\dagger r_k^j{}^\dagger - 
y_k^i y_k^i y_k^j{}^\dagger y_k^j{}^\dagger]
\nonumber\\&-&
2 \left({\alpha' \over 2} \right)^2 
\sum_{k \ne l} 
{2 (p^+)^2 (\varpi_k + \varpi_l)^2 
\over \varpi_k \varpi_l
[(\varpi_k + \varpi_l)^2 - (k+l)^2]} 
[ r_k^i r_l^i r_l^j{}^\dagger r_k^j{}^\dagger - 
y_k^i y_l^i y_k^l{}^\dagger y_k^j{}^\dagger]
\nonumber\\&-&
2 \left({\alpha' \over 2} \right)^2 
\sum_{k \ne l} 
{2 (p^+)^2 (\varpi_k - \varpi_l)^2 
\over \varpi_k \varpi_l
[(\varpi_k - \varpi_l)^2 - (k-l)^2]} 
[ r_k^i r_l^j r_l^i{}^\dagger r_k^j{}^\dagger - 
y_k^i y_l^j y_l^i{}^\dagger y_k^j{}^\dagger]
\hspace{2.5em}
\end{eqnarray}
Expressions 
(\ref{eq:terms from X-ss X-tt X4})-(\ref{eq:terms from X-dot X-dot}) 
appear in ${1 \over p^+} \delta H$ with an overall prefactor of 
\begin{eqnarray}
\label{eq:overall prefactor}
{1 \over 4 \pi \alpha'} \cdot 2 \pi \cdot {1 \over R^2} \cdot {1 \over p^+}
= {1 \over 2 \alpha' R^2 p^+} 
\end{eqnarray}
and we find 
\begin{eqnarray}
\label{eq:delta H}
{1\over p^+} \delta H &=&
{\alpha' \over 4 R^2 (p^+)^3} 
\sum_{i , j} 
\sum_{k}
{k^2 (p^+)^2 \over \varpi_k^2 }
(y^i_k y^i_k y^j_k{}^\dagger y^j_k{}^\dagger 
-r^i_k r^i_k r^j_k{}^\dagger r^j_k{}^\dagger) 
\nonumber\\&+&
{\alpha' \over 4 R^2 (p^+)^3} 
\sum_{i,j} 
\sum_{k \ne l}
{-2 k l (p^+)^2 \over \varpi_k \varpi_l }
(y^i_k y^i_l{}^\dagger y^j_l y^j_k{}^\dagger 
-r^i_k r^i_l{}^\dagger r^j_l r^j_k{}^\dagger) 
\nonumber\\&+&
{\alpha' \over 4 R^2 (p^+)^3} 
\sum_{i,j} 
\sum_{k \ne l} 
{2 k l (p^+)^2 \over \varpi_k \varpi_l }
(y^i_k y^j_k{}^\dagger y^i_l y^j_l{}^\dagger 
-r^i_k r^j_k{}^\dagger r^i_l r^j_l{}^\dagger) 
\nonumber\\&+&
... 
\end{eqnarray}
The ``...'' 
stands for terms not of the form 
$a a^\dagger a a^\dagger$, 
as well as terms with more than two distinct worldsheet momenta; 
we are also dropping corrections which are higher order 
in $1/R^2$ and $1/p^+$. 
The second and third lines of 
(\ref{eq:delta H}) cancel if $i = j$.

In deriving 
(\ref{eq:terms from X-ss X-tt X4})-(\ref{eq:terms from X-dot X-dot}), 
we have not been careful about the ordering of oscillators. 
This means that we may have overlooked some terms 
which involve commutators $[y^i_m , y^i_m{}^\dagger] = 1$. 
The only terms in (\ref{eq:delta H}) 
where this could happen come from the first line. 
This means we could be possibly neglecting 
\begin{eqnarray}
\label{eq:st:P+:correction:relevant:missing}
- \delta' \PP_+ 
= 
\left( {\alpha' \over 4 R^2 (p^+)^3} \right) 
\zeta \sum_{i} \sum_{k} {k^2 (p^+)^2 \over \varpi_k^2} 
(N_k^y{}^i - N_k^r{}^i) 
\end{eqnarray}
If we were to keep track of the ordering of oscillators, 
we would find $\zeta=1$. 
However, we have not analyzed the fermionic side, 
which can also produce similar terms. 

Finally, we compare the results of this Appendix with 
what we found in Section \ref{section:worldsheet}. 
We will only look at the $y$-oscillators. 
The difference between $p^+$ and $\eta$ is 
\begin{eqnarray}
\label{eq:p+ vs eta}
p^+ = 
\eta 
\left[ 
1 + {1\over 2 R^2} \sum_{i;n} {N^{yi}_n \over \varpi_n}
\right]
\end{eqnarray}
so the frequencies in the two approaches are related as 
\begin{eqnarray}
\label{eq:w/s frequencies}
\varpi_m = w_m 
\left[ 
1 + {\eta^2\over 2 R^2 w_m^2} \sum_{i;n} {N^{yi}_n \over w_n}
+ \OO(1/R^4)
\right]
\end{eqnarray}
Expressions 
(\ref{eq:delta H}) and (\ref{eq:st:P+:correction:relevant:missing}) 
then change trivially as 
$\varpi_n \to w_n$, $p^+ \to \eta$ at this 
order in $1/R^2$, while 
\begin{eqnarray}
\label{eq:H0:p+ vs eta}
{1\over p^+} \sum_{i;n} \varpi_n N^{yi}_n 
= 
{1\over \eta} \sum_{i;n} w_n N^{yi}_n 
- 
{1\over 2 R^2 \eta} \sum_{i,j;m,n} 
{n^2 N^{yi}_n N^{yj}_m \over w_m w_n} 
+ \OO(1/R^4)
\end{eqnarray}
Together, 
(\ref{eq:delta H}) and (\ref{eq:H0:p+ vs eta}) 
reproduce 
the sum of 
(\ref{h0o}), (\ref{eq:adrei:diag}) and (\ref{eq:adrei:off-diag}).


\section{\NNN=4 SYM}
\label{section:feynman rules}

Here, we give some details of the \NNN=4 SYM 
needed for the order $g_{\rm Y\!M}^0$ (tree) and $g_{\rm Y\!M}^2$ (one-loop level) 
calculations of Section 
(\ref{section:anomalous dims}). 
First we write down the \NNN=4 SYM action in terms 
of the fields we will be dealing with. 
When SUSY is broken down to \NNN=1, 
things much more cumbersome, 
so from the very beginning we use the \NNN=4 
Lagrangian \cite{Ryzhov}(A.12), 
\begin{eqnarray}
\label{eq:lagrangian:su3 form}
\LL &=& \mbox{$1\over g_{\rm Y\!M}^2$ } \tr 
\left\{ 
- \mbox{$1\over 4$} F_{\mu\nu} F^{\mu\nu} 
+ i \lambda \sigma^\mu D_\mu \bar \lambda 
+ i \psi_j \sigma^\mu D_\mu \bar \psi^j 
+ D_\mu z_j D^\mu \bar z^j 
\right. \\ \nonumber &&\quad~~ \left. 
+ i \sqrt2 [ \lambda , \psi_j ] \bar z^j 
- \mbox{$i\over\sqrt2$} \e^{jkl} [ \psi_j ,  \psi_k ] z_l 
+ i \sqrt2 [ \bar \lambda , \bar \psi^j ] z_j 
- \mbox{$i\over\sqrt2$} \e_{jkl} [ \bar \psi^j , \bar \psi^k ] \bar z^l 
\right. \\ \nonumber &&\quad~~ \left. 
+ [ z_j , z_k ] [ \bar z^j , \bar z^k ] 
- \half [ z_j , \bar z^j ] [ z_k , \bar z^k ] 
\right\}
\end{eqnarray}
we leave the fields $z_1$, $\bar z^1$ as they are, 
and substitute 
\begin{equation}
\label{eq:z vs phi}
     z_j = \mbox{$1\over\sqrt{2}$} \left( \phi_j + i \phi_{j+3} \right) , 
\quad
\bar z^j = \mbox{$1\over\sqrt{2}$} \left( \phi_j - i \phi_{j+3} \right)  , 
\quad
\mbox{$j=2$ and 3.}
\end{equation}
The rest of the fields (gauge bosons and fermions) remain unchanged, 
and (\ref{eq:lagrangian:su3 form}) becomes 
\begin{eqnarray}
\label{eq:lagrangian:mixed form:def}
\LL = \LL_0 + \LL_1 + \LL_2 + \LL_{\rm other}
\end{eqnarray}
where
\begin{eqnarray}
\label{eq:lagrangian:mixed form:propagators}
\LL_0 &=& \mbox{$1\over g_{\rm Y\!M}^2$ } \tr 
\left\{ 
(\partial_\mu z_1) (\partial^\mu \bar z_1) 
+ 
\sum_k
\half 
(\partial_\mu \phi_k) (\partial^\mu \phi_k) 
\right\}
\end{eqnarray}
gives propagators for the scalars; 
\begin{eqnarray}
\label{eq:lagrangian:mixed form:order g}
\LL_1 &=& \mbox{$1\over g_{\rm Y\!M}^2$ } \tr 
\left\{ 
- i A^\mu [ z_1 , \partial_\mu \bar z_1 ] 
- i A^\mu [ \bar z_1 , \partial_\mu z_1 ] 
+ 
\sum_k
(- i A^\mu) [ \phi_k , \partial_\mu \phi_k ] 
\right. \\ \nonumber && \left. 
+ i \sqrt2 \; z_1 
\left( 
[ \bar \lambda , \bar \psi^1 ] - [ \psi_2,  \psi_3 ] 
\right) 
+ i \sqrt2 \; \bar z_1 
\left( 
[ \lambda , \psi_1 ] - [ \bar \psi^2,  \bar \psi^3 ] 
\right) 
\right. \\ \nonumber && \left. 
+ i \phi_2 
\left( 
[ \lambda , \psi_2 ] + [ \bar \lambda , \bar \psi^2 ] 
- [ \psi_3,  \psi_1 ] - [ \bar \psi^3,  \bar \psi^1 ] 
\right) 
+ \phi_5 
\left( 
[ \lambda , \psi_2 ] - [ \bar \lambda , \bar \psi^2 ] 
+ [ \psi_3,  \psi_1 ] - [ \bar \psi^3,  \bar \psi^1 ] 
\right) 
\hspace{-3em}
\right. \\ \nonumber && \left. 
+ i \phi_3 
\left( 
[ \lambda , \psi_3 ] + [ \bar \lambda , \bar \psi^3 ] 
- [ \psi_1,  \psi_2 ] - [ \bar \psi^1,  \bar \psi^2 ] 
\right) 
+ \phi_6 
\left( 
[ \lambda , \psi_3 ] - [ \bar \lambda , \bar \psi^3 ] 
+ [ \psi_1,  \psi_2 ] - [ \bar \psi^1,  \bar \psi^2 ] 
\right) 
\right\}
\hspace{-3em}
\end{eqnarray}
gives 3-field vertices; 
and 
\begin{eqnarray}
\label{eq:lagrangian:mixed form:order g^2}
\LL_2 &=& \mbox{$1\over g_{\rm Y\!M}^2$ } \tr 
\left\{ 
- [ A_\mu , z_1 ] [ A^\mu , \bar z_1 ]  
- \sum_k \half [ A_\mu , \phi_k ] [ A^\mu , \phi_k ]  
\right. \\ \nonumber &&\quad~~ \left. 
- \half [ z_1 , \bar z_1 ] [ z_1 , \bar z_1 ] 
+ \sum_k [ z_1 , \phi_k ] [ \bar z_1 , \phi_k ] 
+ \sum_{k > l} \half [ \phi_k , \phi_l ] [ \phi_k , \phi_l ] 
\right\}
\end{eqnarray}
contains 4-field interactions. 
Finally, 
\begin{eqnarray}
\label{eq:lagrangian:mixed form:other}
\LL_{\rm other} &=& \mbox{$1\over g_{\rm Y\!M}^2$ } \tr 
\left\{ 
- \mbox{$1\over 4$} F_{\mu\nu} F^{\mu\nu} 
+ i \lambda \sigma^\mu D_\mu \bar \lambda 
+ i \psi_j \sigma^\mu D_\mu \bar \psi^j 
\right\}
\end{eqnarray}
gives propagators for the gauge bosons and the fermions 
and their interactions with each other (at order $\OO(g_{\rm YM}^2)$ 
these do not contribute to the diagrams we care about, and neither 
do the ghost terms). 
The Lagrangian (\ref{eq:lagrangian:mixed form:def}) 
has a leftover $SO(4)$ symmetry rotating the $\phi$-s.

\begin{figure}
{\begin{center}
\epsfig{width=4in, file=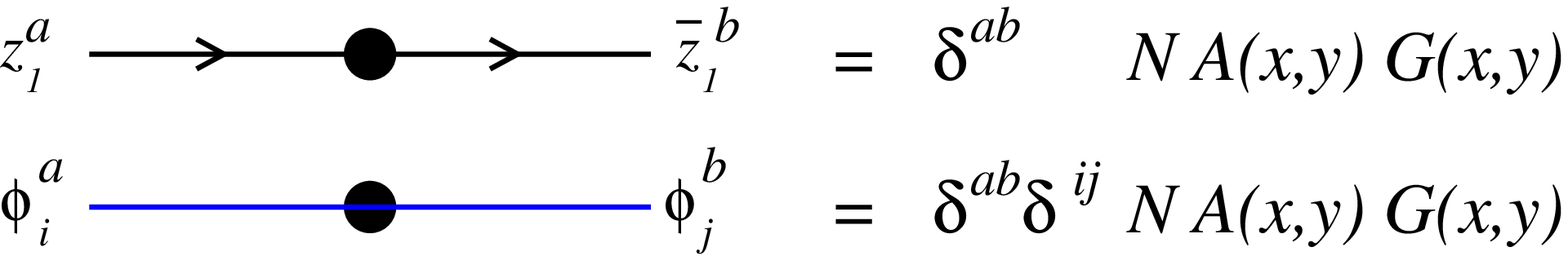, angle=0}\quad
\end{center}}
\vskip -0.2in
\caption{%
Order $g_{\rm Y\!M}^2$ corrections to scalar propagators 
consist of a gauge boson exchange and a fermion loop. 
\label {fig:propagator}
}%
\end {figure}

Feynman rules for the Lagrangian (\ref{eq:lagrangian:mixed form:def}) 
are somewhat awkward, but the tree and one-loop diagrams 
which involve only the scalars can be packaged in a convenient way. 
First, $\OO(g_{\rm Y\!M}^2)$ corrections to the 
scalar propagators are diagonal in color indices, 
see Figure \ref{fig:propagator}. 
Fermion loops cancel in $\langle \phi_2^a(x) \phi_5^b(y) \rangle_{g_{\rm Y\!M}^2}$ 
because of the way the signs work out in 
(\ref{eq:lagrangian:mixed form:order g}).

\begin{figure}
{\begin{center}
\epsfig{width=6.2in, file=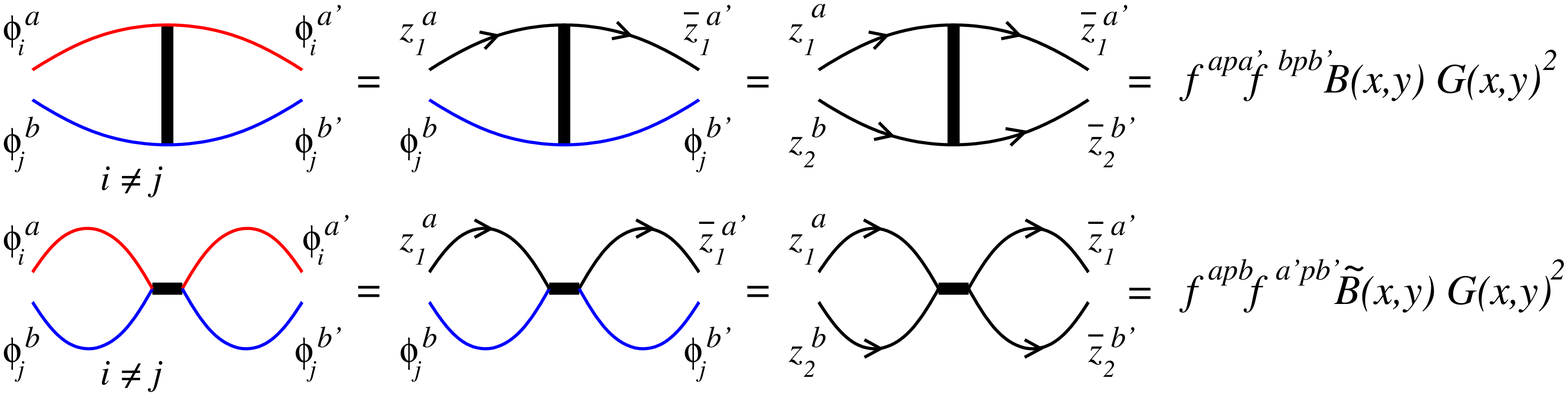, angle=0}\quad
\end{center}}
\vskip -0.2in
\caption{%
Order $g_{\rm Y\!M}^2$ corrections to two point functions of operators of the form 
$\tr z ... \phi^1 z ... \phi^2$: four-field irreducible blocks. 
When scalars $\phi^i$ are involved, 
the diagrams above represent the net contribution 
of all contributing Feynman diagrams, packaged in 
a way to mimic the \NN=1 component fields 
Feynman diagrams. 
(Thick lines would correspond to exchanges of auxiliary 
fields $F_i$ and $D$ in the \NN=1 formulation.) 
Diagrams with $z_2$ are given for comparison only. 
There are similar diagrams with 
one or both $z$-lines running in the opposite direction. 
\label {fig:four-point irreducible blocks}
}%
\end {figure}

\begin{figure}
{\begin{center}
\epsfig{width=6.2in, file=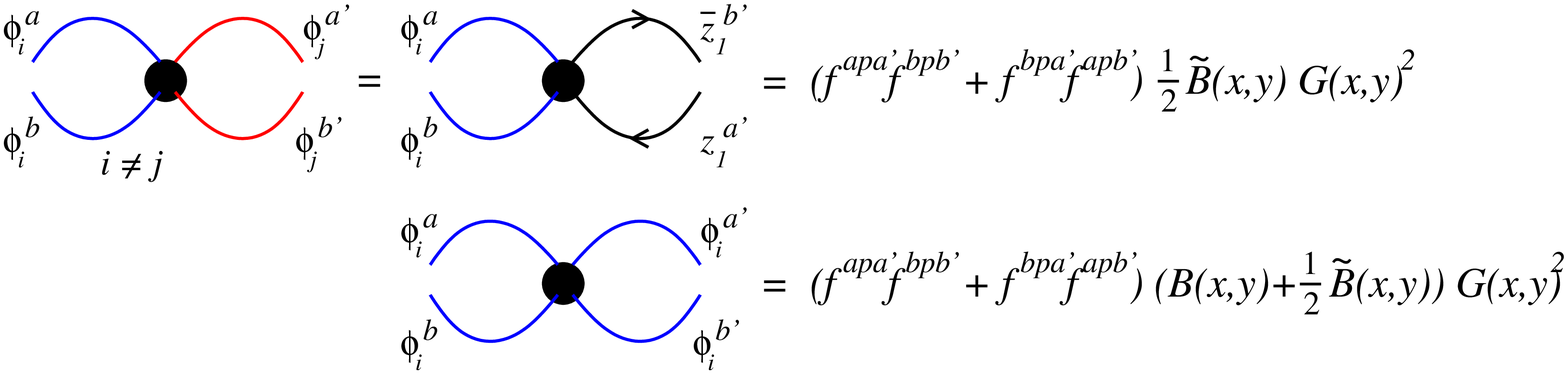, angle=0}\quad
\end{center}}
\vskip -0.2in
\caption{%
Order $g^2$ corrections to two point functions of operators of the form 
$\tr z ... \phi^1 z ... \phi^1$: 
four-field irreducible blocks.
Thick lines correspond to exchanges if the gauge boson and auxiliary 
fields $F_i$ and $D$ in the \NN=1 formulation. 
The diagrams above represent the net contribution 
of all contributing Feynman diagrams, packaged in 
a way to mimic the \NN=1 component fields 
Feynman diagrams. 
\label {fig:four-point irreducible blocks:same}
}%
\end {figure}

Corrections to the 4-point irreducible blocks 
are more involved, but they can be related to the 
corresponding diagrams involving only $z$-s and $\bar z$-s. 
By comparing two-point functions of the protected 
operators in the [0,2,0] of $SU(4)$ 
written on the one hand in terms 
of $\phi$-fields, and on the other hand in terms of $z$-s and $\bar z$-s, 
we get the diagrams shown in 
Figure \ref{fig:four-point irreducible blocks}. 
Comparison of two-point functions of the Konishi scalar 
$\sum_{k=1}^6 \tr \phi^k \phi^k = \sum_{k=1}^3 \tr z^k \bar z^k$ 
produce the relations listed in 
Figure \ref{fig:four-point irreducible blocks:same}.

The ``$D$-term'' contributions $A$ and $B$, 
and the four-field interaction ``$F$-term'' $\tilde B$ are 
defined by Figures \ref{fig:four-point irreducible blocks} 
and \ref{fig:four-point irreducible blocks:same}. 
As in \cite{Ryzhov} \cite{CFHMMPS}, the 
$A$ and $B$ are not separately gauge invariant. 
These must appear as the gauge invariant combination $2A+B$, 
which vanishes 
in the \NNN=4 theory. So one only has to 
look at ``$F$-term'' contributions, which are all proportional to 
\begin{eqnarray}
\label{eq:b-tilde}
\gamma \equiv 
\half \tilde B (x,0) N = 
- {g_{\rm Y\!M}^2 N \over 4 \pi^2 } 
\log x^2 \mu^2 
\equiv 
- \beta 
\log x^2 \mu^2 
\end{eqnarray}
computed for example in \cite{BMN,Ryzhov}. 
In this paper, we are using the conventions of \cite{BMN}; 
in the Lagrangian (\ref{eq:lagrangian:su3 form}) we have 
$g_{\rm Y\!M}^2 = 2 \pi g$.

We only have to consider planar diagrams 
since we are interested in the leading large $N$ behavior. 
Put differently, 
\begin{eqnarray}
\label{eq:traces-leading}
\tr \! \left[ t^{a_1} ... t^{a_k} \right]
~
\tr \! \left[ t^{a_k} ... t^{a_1} \right] 
= \left( {N \over 2} \right)^k \left[ 1 + \OO(1/N^2) \right]
\end{eqnarray}
and $SU(N)$ traces of all other permutations of the generators 
(other than cyclic) are suppressed by $1/N^2$. 
To see this, one can use the ``trace merging formula''
\begin{equation}
\label{eq:merging traces}
2 \left( \tr A t^c \right) \left( \tr B t^c \right) 
= 
\tr A B - 
\mbox{$1\over N$} \left( \tr A \right) \left( \tr B \right) 
\end{equation}
valid when $t^c$ are $SU(N)$ generators in the fundamental 
representation.

At one loop, 
all but the nearest neighbor interactions are suppressed. 
The relevant contributions in 
Figure \ref{fig:four-point irreducible blocks} 
have the form 
\begin{eqnarray}
\label{eq:one-loop combinatorial factor}
\XII{a}{a'}{b}{b'}{c_1}{c_J}
&=& 
\tr \! \left[ t^a t^b t^{c_1} ... t^{c_J} \right]
~
\tr \! \left[ t^{c_J} ... t^{c_1} t^{b'} t^{a'} \right] 
~
f^{abp} f^{a'b'p}
\nonumber\\
&=& \half \left( \half N \right)^{J-1} 
\tr \left( t^a t^b t^{b'} t^{a'} \right) f^{abp} f^{a'b'p}
\left[ 1 + \OO(1/N^2) \right]
\nonumber\\
&=& \half \left( \half N \right)^{J-1} 
\tr \left( t^a [t^p, t^a] t^{b'} [t^p , t^{b'}] \right) 
\left[ 1 + \OO(1/N^2) \right]
\nonumber\\
&=& \left( \half N \right)^{J+3}
\left[ 1 + \OO(1/N^2) \right]
\end{eqnarray}
The difference between the orderings $(ab)$ and $(ba)$ 
in (\ref{eq:one-loop combinatorial factor}) 
is a minus sign, 
\begin{eqnarray}
\label{eq:orderings}
\XII{a}{a'}{b}{b'}{c_1}{c_J}
= - 
\XII{b}{a'}{a}{b'}{c_1}{c_J}
\end{eqnarray}
Diagrams shown in the first two lines of 
Figure \ref{fig:four-point irreducible blocks:same}
have the form 
\begin{eqnarray}
\label{eq:one-loop combinatorial factor:blob}
\blobII{a}{a'}{b}{b'}{c_1}{c_J}
&=& 
\half [ f^{aa'p} f^{bb'p} + f^{ab'p} f^{ba'p} ] 
~ 
\tr \! \left[ t^a t^b t^{c_1} ... t^{c_J} \right]
~
\tr \! \left[ t^{c_J} ... t^{c_1} t^{b'} t^{a'} \right] 
\hspace{-2em}
\nonumber\\
&=& \half \left( \half N \right)^{J+3}
\left[ 1 + \OO(1/N^2) \right]
\end{eqnarray}
Only one of the two $f f$-terms contributes at this order; 
the other one is suppressed by at least $1/N^3$. 
The contribution (\ref{eq:one-loop combinatorial factor:blob}) 
is insensitive to $a \lra b$. 
The contributions (\ref{eq:one-loop combinatorial factor}) 
and (\ref{eq:one-loop combinatorial factor:blob}) 
come with a numerical prefactor of 
\begin{eqnarray}
\label{eq:prefactor:def}
{2 \over N} G(x,0)^{J+2} 
\gamma 
\end{eqnarray}
with $\gamma = - \beta \log x^2 \mu^2$ defined in (\ref{eq:b-tilde}).

To summarize, the tree level correlators are 
\begin{eqnarray}
\label{eq:tree diagrams}
\matrix{\epsfig{height=1.1in, file=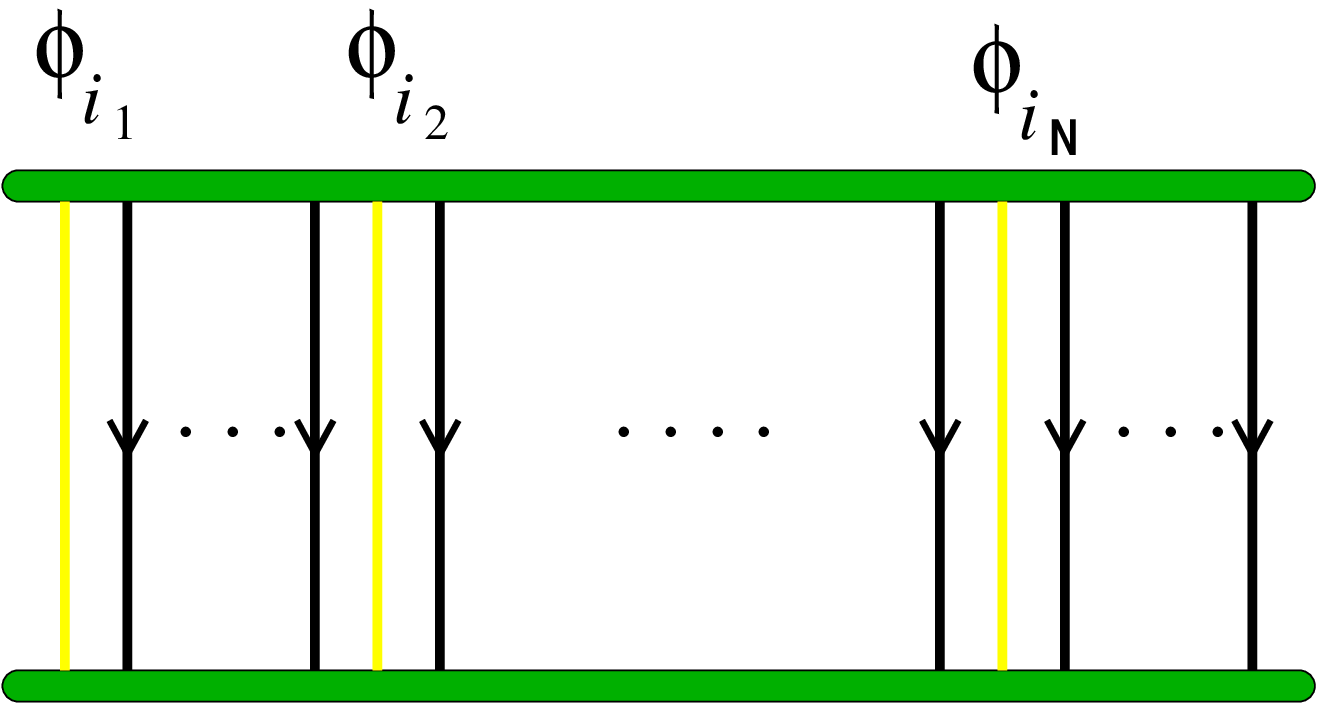}\quad}
= 
(\half G N)^{J+\N} 
\end{eqnarray}
and the relevant 
one-loop contributions 
can be schematically represented as 
\begin{eqnarray}
\label{eq:one-loop diagrams:one phi}
\matrix{\epsfig{height=0.7in, file=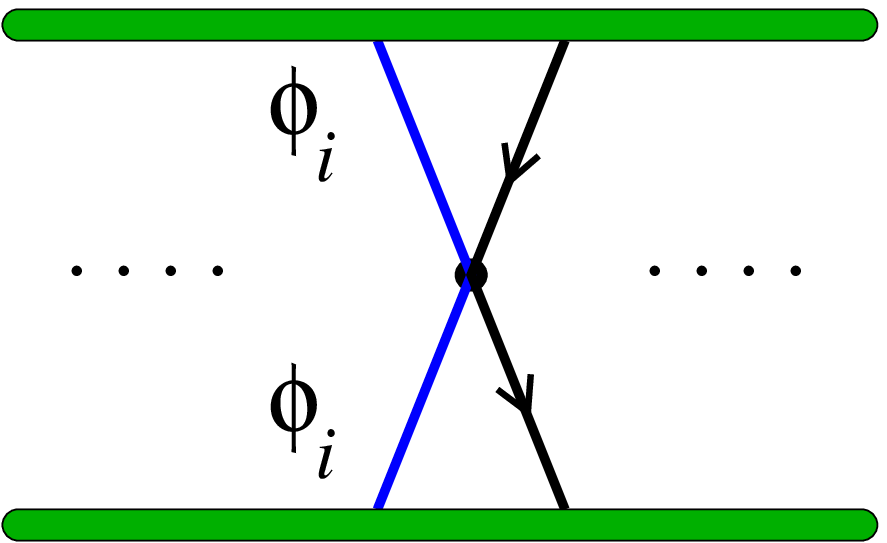}}
= - 
\matrix{\epsfig{height=0.7in, file=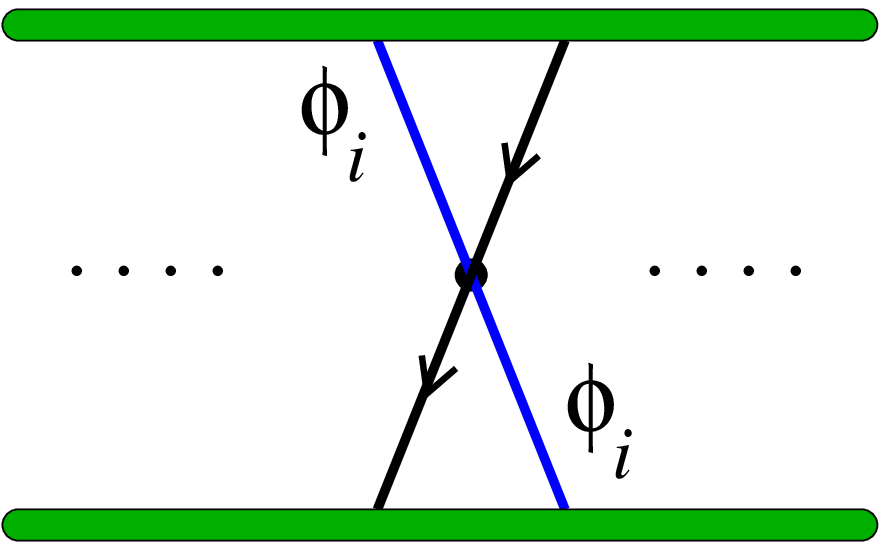}}
&=& 
\gamma \times (\half G N)^{J+\N} 
\end{eqnarray}
when only one $\phi$ is involved in the interaction, 
and  
\begin{eqnarray}
\label{eq:one-loop diagrams:two phi:distinct}
\matrix{\epsfig{height=0.7in, file=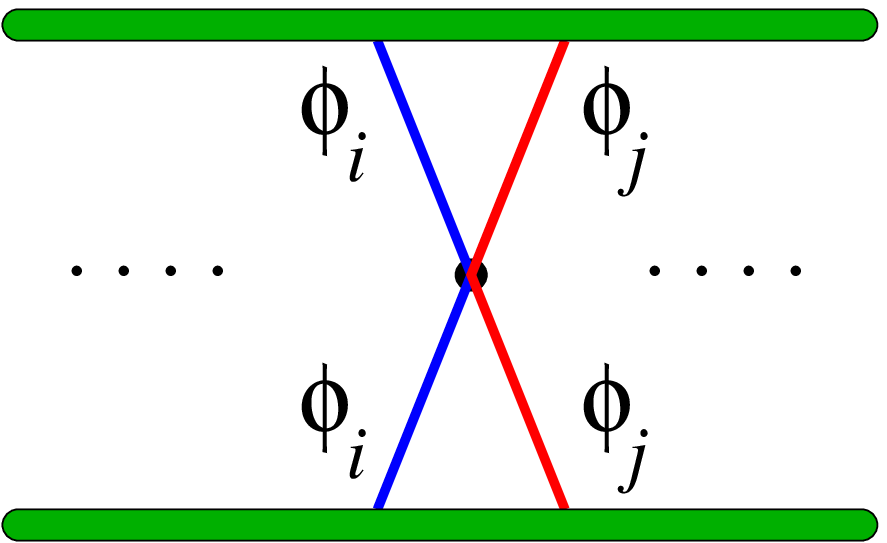}}
= - 
\matrix{\epsfig{height=0.7in, file=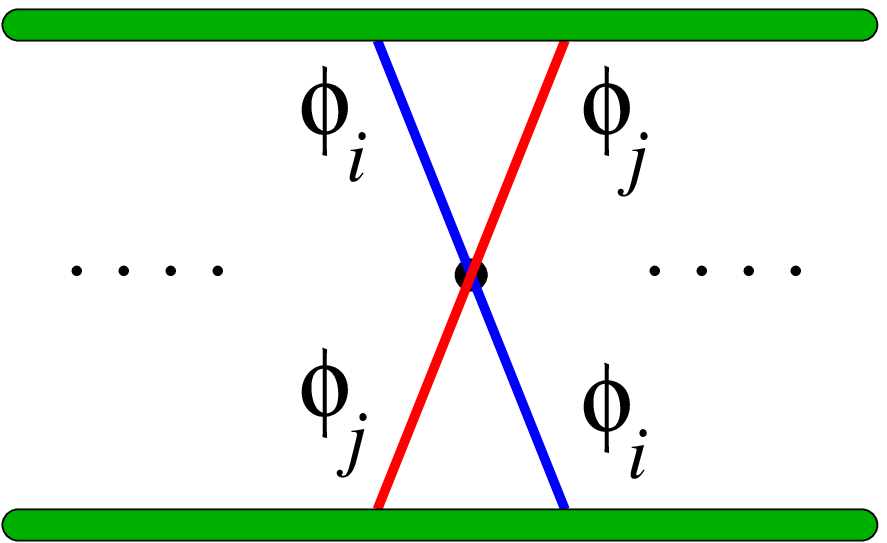}}
&=& 
\gamma \times (\half G N)^{J+\N} 
,\quad i \ne j
\end{eqnarray}
when two distinct $\phi$ within either trace interact. 
Furthermore, we have 
\begin{eqnarray}
\label{eq:one-loop diagrams:two phi:same}
\matrix{\epsfig{height=0.7in, file=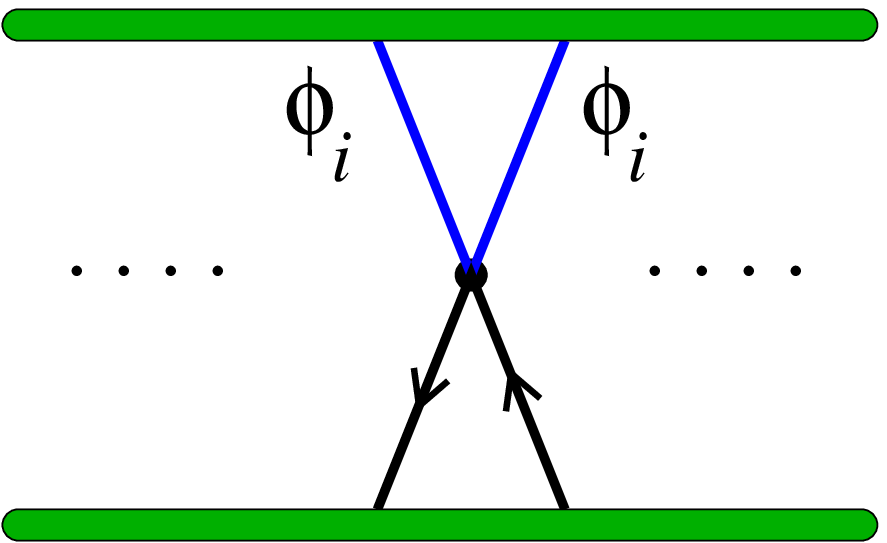}}
= 
\matrix{\epsfig{height=0.7in, file=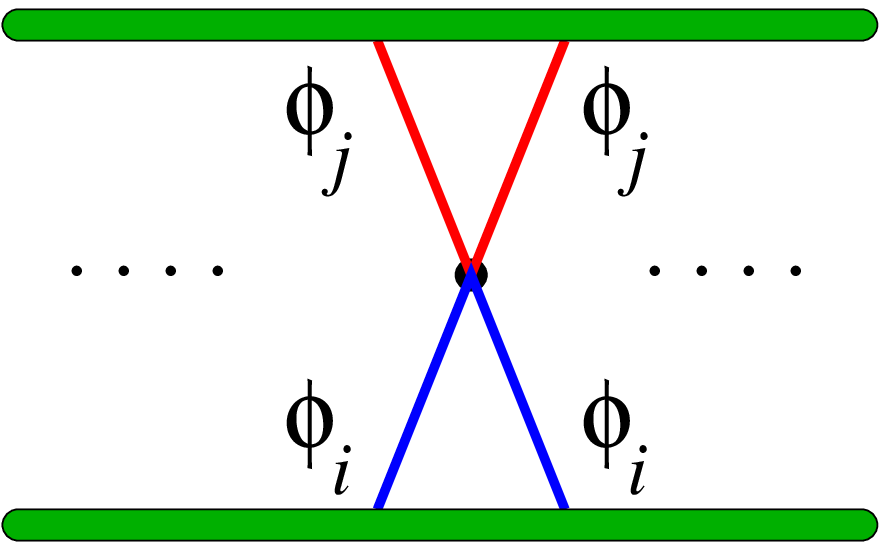}}
= 
\matrix{\epsfig{height=0.7in, file=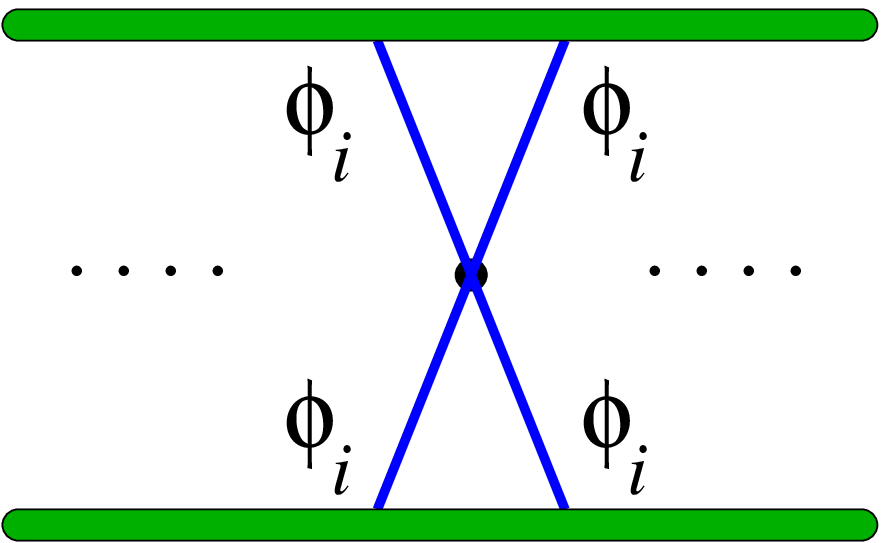}}
&=& 
{1\over2} \, \gamma \times (\half G N)^{J+\N} 
,\quad i \ne j
\nonumber\\
\end{eqnarray}
Finally, the diagrams which involve a $z z \bar z \bar z$ vertex 
can be read off from 
(\ref{eq:one-loop diagrams:two phi:distinct}) 
and (\ref{eq:one-loop diagrams:two phi:same}) 
by expanding the $z$ and $\bar z$ participating in the vertex 
in terms of the two remaining $\phi$'s, 
\begin{eqnarray}
\label{eq:one-loop diagrams:z-zbar}
\matrix{\epsfig{height=0.7in, file=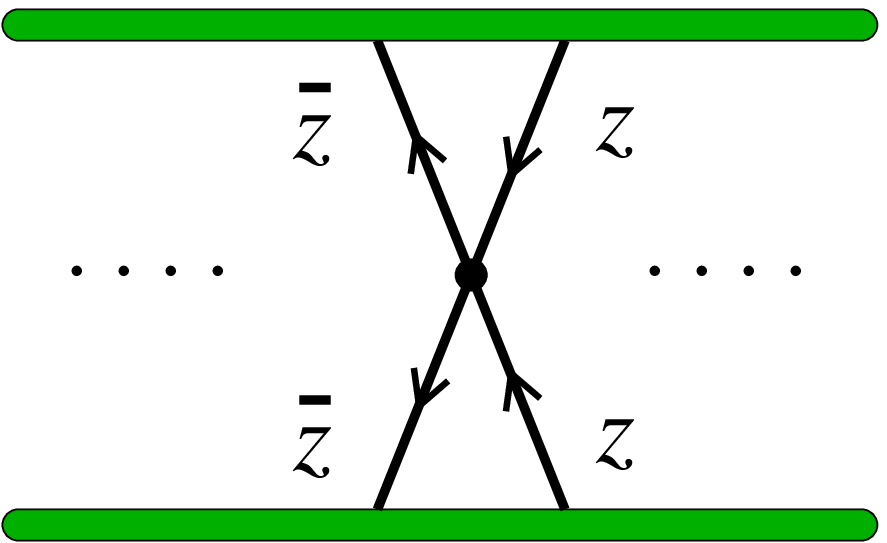}}
= 
- {1\over2} \, \gamma \times (\half G N)^{J+\N} 
, \hspace{2em}
\matrix{\epsfig{height=0.7in, file=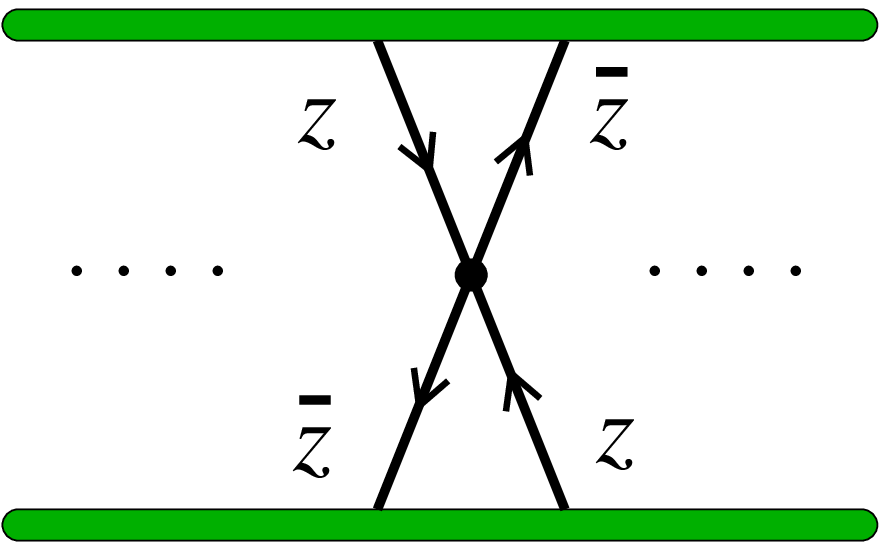}}
&=& 
{3\over2} \, \gamma \times (\half G N)^{J+\N} 
\nonumber\\
\end{eqnarray}
%
In the results 
(\ref{eq:tree diagrams})-(\ref{eq:one-loop diagrams:z-zbar}), 
we dropped terms suppressed by $1/N^2$.

\section{Equality of matrix elements: generic states}
\label{section:equalitygeneric}
In this appendix we will complete matching the matrix
elements of the light cone Hamiltonian between the two sides
of the AdS/CFT correspondence.
In section \ref{section:equalitysimple} we matched 
matrix elements for a subset of states.
There we considered states with all excited modes having distinct SO(4) indices $i_k$,
and no mode excited more than once.
We will now consider states with some $i_k$ being equal.
We initially restrict to the case with no modes excited
more than once, $N^i_n \le 1$, but will eventually generalize to most
general case.

In contrast with section \ref{section:equalitysimple}, $\Or_*$ no longer vanishes.
In addition to (\ref{del1}), we now have to consider off-diagonal elements between the states
\beq
\label{i2}
  |\Y\rangle
        =  y_{n_1}^{i_1}{}^\dagger  \ldots  y_m^i{}^\dagger  
        y_n^i{}^\dagger  |\eta \rangle, \qquad m \neq n,
\eeq
and
\beq
\label{f2}
  |\Y'\rangle
      =  y_{n_1}^{i_1}{}^\dagger  \ldots  
   y_m^j{}^\dagger  y_n^j{}^\dagger  |\eta \rangle, \qquad m \neq n,
\eeq
which are given by
\beq
\label{del2}
  \langle \Y|  H_1^{OD} |\Y' \rangle  =  
      {1 \over J} \left({ R^2 \over J}\right)^2 m n \sqrt{N^i_m N^i_n N^j_m{}' N^j_n{}'}.
\eeq
There is also an off-diagonal element given by (\ref{del1}),
but we have analyzed all diagrams contributing to it in 
section \ref{section:equalitysimple}.
Let us briefly explain why this is the case.
Consider $\Or(g^0)$ part of the contributing two-point function,
which we denote by $\langle \X {\bar \X}'\rangle_{g^0}$.
$\langle {\tilde \X} {\bar \X}_*'\rangle_{g^0}$ and
$\langle \X_*' {\bar {\tilde \X}} \rangle_{g^0}$
vanish, as there are no contributing interaction-free diagrams.
Although $\langle \X_* {\bar \X'}_*\rangle_{g^0}$  has 
nonvanishing terms, they are $\Or(1/J^2)$.
This is because $\X_*$ is itself $\Or(1/J)$ compared to ${\tilde \Or}$,
and an additional factor of $1/J$ will
appear because phases in $\X_*$ and $\X_*'$ do not
match exactly.
Similar conclusions can be made about $\Or(g)$ correlator 
$\langle {\tilde \X} {\bar \X'}\rangle_{g^1}$.

Let us compute the off-diagonal element (\ref{del2}) in the gauge theory.
The only contribution to $\langle \Y(x) {\bar \Y'}(0)\rangle_{g^0}$
comes from 
\beq
\label{intfree3}
 \langle \Y_*(x) {\bar \Y'}_*(0)\rangle_{g^0}=
{1 \over \Omega} \sum 
\III{{\check \phi}^i_m}{{\check \phi}^j_{m}}{{\check \phi}^i_n}{{\check \phi}^j_{n}}
 { \phi^{i_k}_{n_k}}{ \phi^{i_k}_{n_k}}={1 \over J},
\eeq
where the sum runs over all configurations of fields. 
No $\Or(g^0)$ diagrams appear in $\langle {\tilde \Y} {\bar \Y'}_* \rangle_{g^0}$,
$\langle \Y_* {\bar {\tilde \Y'}} \rangle_{g^0}$ and 
$\langle {\tilde \Y} {\bar {\tilde \Y'}} \rangle_{g^0}$.
Hence we have
\beq
\label{iijjod}
   {\bf T}^{(1)}_{\Y\Y'}=1.
\eeq

Computation of $\F_{\Y\Y'}$ is more involved.
Possible contributions are
\beqa
\label{cr3}
\langle {\tilde \Y}(x) {\bar {\tilde \Y'}}(0) \rangle_{g^1}&=&
   {1 \over \Omega} {\sum} {\XIIm{\phi^i_m}{\phi^j_{m'}}{\phi^i_n}{\phi^j_{n'}}
           {z}{\bar z}{\phi^{i_k}_{n_k}}{\phi^{i_k}_{n_k}}} {+}
   (m {\lra} n) {+} (m' {\lra} n') {+}(m {\lra} n, m' {\lra} n') \\ \nonumber \\ \nonumber
&=& { \gamma \over 2 J} (q_{n-m}+q_{n-m}^*+2),  
\eeqa 
which holds both for $m'=n, n'=m$ (off-diagonal) and
$m'=m, n'=n$ (diagonal),
\beqa
 \label{cr4}
{-}\langle {\tilde \Y}(x) {\bar \Y'}_*(0) \rangle_{g^1}&=&
 {-}{1 \over \Omega} {\sum}  \XIIm{\phi^i_m}{z}{\phi^i_n}{\bar z}
           {z}{\bar z}{\phi^{i_k}_{n_k}}{\phi^{i_k}_{n_k}} {+}
 (m {\lra} n) {+} (z {\lra} {\bar z}) {+}(m {\lra} n, z {\lra} {\bar z})
       \\ \nonumber \\ \nonumber
&=& -{ \gamma \over 2 J} (q_m+q_m^*+q_n+q_n^*),
\eeqa  
similar contribution from $ \langle \Y_*(x) {\bar \Y'}_*(0)\rangle_{g^1}$,
and
\beqa
\label{cr5}
&&\langle \Y_*(x) {\bar \Y'}_*(0) \rangle_{g^1}=
{1 \over \Omega} {\sum}  \IIX{z}{\bar z}{\;\;\bar z_{m{+}n}}{\;\;z_{m{+}n}}{\phi^{i_k}_{n_k}}{\bar z}{z}{\phi^{i_k}_{n_k}}{+}
          (\phi^{i_k}_{n_k} {\lra} z){+}(\phi^{i_k}_{n_k} {\lra} {\bar z}) \\ \nonumber  
&&\qquad \qquad \qquad \qquad \qquad \qquad \qquad \qquad \qquad
{+}(\phi^{i_k}_{n_k} {\lra} z,\phi^{i_k}_{n_k} {\lra} {\bar z})
\\ \nonumber 
  && \qquad \qquad {+} {1 \over \Omega} \sum \XIIm{z}{\bar z}{\bar z_{m{+}n}}{z_{m{+}n}}
           {z}{\bar z}{\phi^{i_k}_{n_k}}{\phi^{i_k}_{n_k}}
   {+}(z{\lra}{\bar z_{m{+}n}}){+}({\bar z}\lra z_{m{+}n}){+}
        (z{\lra}{\bar z_{m{+}n}},{\bar z}\lra z_{m{+}n})
 \\ \nonumber \\ \nonumber 
 && \qquad = {\gamma \over J} \left( -\sum_{p:\, n_p \neq m,n} (q_{n_p}+q_{n_p}^*-2)
         -{1 \over 2} (q_{m+n}+q_{m+n}^*) + 3 \right). 
\eeqa
(Recall that the subscript in ${\bar z}_{n+m}$ stands for the phase
$q_{n+m}^{a_{\bar z}}$ which depends on the position of the $\bar z$ in the string
of operators.)
Combining (\ref{iijjod})--(\ref{cr5}) we have
\beq
\label{del2a}
  [{\bf T}^{-1} \F]_{\Y\Y'}= -{\beta \over 2 J} (q_{m+n}-q_{m-n}+c.c.)=
             {1 \over J} \left({R^2 \over J}\right)^2 m n,
\eeq
which indeed agrees with (\ref{del2}), provided $N^i_n \le 1$.

Let us now turn to the diagonal matrix element.
Part of it was computed in section \ref{section:equalitysimple}
and is given by (\ref{tfdiag}). 
But now there are other contributions both to ${\bf T}^{(1)}_{\Or\Or}$
and to $\F_{\Or\Or}$.
To update the former, we must take into account
\beq
\label{intfree4}
 \langle \Or_*(x) {\bar \Or}_*(0) \rangle_{g^0}= 
  {1 \over \Omega} \sum_{i,(m\neq n)} \sum
    \III{{\check \phi}^i_m}{{\check \phi}^i_{m}}{{\check \phi}^i_n}{{\check \phi}^i_{n}}
                    { \phi^{i_k}_{n_k}}{ \phi^{i_k}_{n_k}}=
     \sum_i {\N_i (\N_i-1) \over 2 J},
\eeq             
and
\beq
\label{intfree5}
 \delta \langle {\tilde \Or}(x) {\bar {\tilde \Or}}(0) \rangle_{g^0}= 
  {1 \over \Omega} \sum_{i,(m\neq n)} \sum
    \III{\phi^i_m}{\phi^i_{n}}{\phi^i_n}{\phi^i_m}
                    { \phi^{i_k}_{n_k}}{ \phi^{i_k}_{n_k}}= 
         -  \sum_i {\N_i (\N_i-1) \over  2 J},
\eeq
which cancels (\ref{intfree4}) to keep ${\bf T}_{\Or\Or}=1$.
The $\Or(g^1)$ correlators related to (\ref{intfree4}) are
given by the sum of (\ref{cr5}) over pairs $(n_k\neq n_l):i_k=i_l$ with
the substitution $m=n_k, n=n_l$: 
\beqa
\label{cr6}
   \langle \Or_*(x) {\bar \Or}_*(0) \rangle_{g^1}&=& 
      {\gamma \over J}\sum_{(n_k{\neq}n_l):i_k{=}i_l}
  \left( 3{-}\sum_{p{\neq}k,l} (q_{n_p}{+}q_{n_p}^*{-}2)
         - {1 \over 2} ( q_{n_k{+}n_l}{+}q_{n_k{+}n_l}^* )
            \right) \\ \nonumber \\ \nonumber
   &=&   {\gamma \over J} \Bigg[ -\left( \sum_i {\N_i (\N_i{-}1) \over 2} \right)
                                \sum_k (q_{n_k}{+}q_{n_k}^*{-}2) \\ \nonumber
    &&\quad+\sum_{(n_k{\neq}n_l):i_k=i_l} \left\{ (q_{n_k}+q_{n_l}-2+c.c.)
                -{1 \over 2} (q_{n_k+n_l}+q_{n_k+n_l}^*) + 3 \right\} \Bigg].
\eeqa
The $\Or(g^1)$ counterpart of (\ref{intfree5}) is
\beq
\label{cr7}
 \delta_1 \langle {\tilde \Or}(x) {\bar {\tilde \Or}}(0) \rangle_{g^1}{=} 
      {\gamma \over J} \bigg[ \left( \sum_i {\N_i (\N_i{-}1) \over 2} \right) 
                \sum_k (q_{n_k}{+}q_{n_k}^*{-}2)
      + 
\!\!
\sum_{(n_k{\neq}n_l):i_k{=}i_l} 
\!\!
( q_{n_k} {+}q_{n_l}{-}2 q_{n_l{-}n_k}{+}c.c.) \bigg].
\eeq
The first term in this expression is a value of the corresponding
interaction-free diagram times the sum of possible phases, while the
second term takes care of overcounted corrections (this technique
for computing $\Or(g^1)$ diagrams was explained in more detail
in section \ref{section:equalitysimple})
There is also a contribution which is a direct analog of (\ref{diagmee})
\beq
\label{cr8}
 \delta_2 \langle {\tilde \Or}(x) {\bar {\tilde \Or}}(0) \rangle_{g^1}{=} 
  { \gamma \over J}  \sum_{(n_k{\neq}n_l):i_k{=}i_l} 
     \left[ (q_{n_k}+q_{n_l}-2+c.c.)+{1 \over 2}(q_{n_k-n_l}+q_{n_k-n_l}^*+2)  \right].
\eeq   
Finally, we should include the sum over pairs in (\ref{cr4}) and
the same term due to 
\beq
\label{cr9}
    -\langle \Y_*(x) {\bar{\tilde  \Y'}}(0){+} {\tilde \Y}(x) {\bar \Y'}_*(0)\rangle_{g^1}=
      - { \gamma \over  J} \sum_{(n_k{\neq}n_l):i_k{=}i_l} (q_{n_k}+q_{n_l}+c.c.),
\eeq
Combining (\ref{cr6})--(\ref{cr9}) we get
\beqa
\label{tfdiag2}
  \delta [{\bf T}^{-1} \F]_{\Or\Or} &=& 
       -{\beta \over 2 J}  \sum_{(n_k{\neq}n_l):i_k{=}i_l} \left[
      3 q_{n_k{-}n_l}{+}q_{n_k{+}n_l}{-}4 q_{n_k}{-}4 q_{n_l}{+}4{+}c.c. \right] \\ \nonumber
   &=&   {-}{1 \over J} \left( {R^2 \over J} \right)^2 \sum_{(n_k{\neq}n_l):i_k{=}i_l} n_k n_l,
\eeqa
which should be added to (\ref{tfdiag}).
In the case of $N^i_m \le 1$, (\ref{tfdiag2}) combined with the
last term in (\ref{tfdiag})
gives
\beq
\label{funnyc}
   -{1 \over J} \left( {R^2 \over J} \right)^2 \left[
       \sum_{(k,l):i_k{\neq}i_l} n_k n_l + \sum_{(n_k{\neq}n_l):i_k{=}i_l} n_k n_l \right]
    = {1 \over 2 J} \left( {R^2 \over J} \right)^2 \sum_k n_k^2,
\eeq
where we used the level matching condition.
Hence we again reproduce (\ref{hdiagx}).

Our last step will be generalization to the case of unconstrained $N^i_n$.
To see how (\ref{del1a}) is modified recall that all contributing
correlators should be divided by
\beq
\label{odf1}
   \sqrt{ N^i_m! N^j_n! N^i_n! N^j_m! N^i_m{}'! N^j_n{}'! N^i_n{}'! N^j_m{}'!\ldots}
\eeq
where $\ldots$ stands for other $N^{i_k}_{n_k}$ which will be cancelled
by the number of possible contractions, just as they are cancelled
in non-interacting diagrams to produce $T_{\Or\Or}=1+\Or(1/J)$.
On the other hand, the combinatorial factor that multiplies all the
correlators contributing to (\ref{del1a}) is
\beq
\label{odf2}
    N^i_m! N^j_n! N^i_n{}'! N^j_m{}'! \ldots
\eeq
The ratio of (\ref{odf2}) and (\ref{odf1}) is precisely
the factor $\sqrt{N^i_m N^j_n N^i_n{}' N^j_m{}'}$ which appears in (\ref{del1a}).
The combinatorial factor in (\ref{del2}) can be restored in the similar manner.

In addition to (\ref{del1}) and (\ref{del2}) we also need to consider
off-diagonal matrix elements between the states
\beq
\label{i3}
  |{\cal Z}\rangle
        =  y_{n_1}^{i_1}{}^\dagger  \ldots  y_n^i{}^\dagger  
        y_n^i{}^\dagger  |\eta \rangle, 
\eeq
and
\beq
\label{f3}
  |{\cal Z}'\rangle
      =  y_{n_1}^{i_1}{}^\dagger  \ldots  
   y_n^j{}^\dagger  y_n^j{}^\dagger  |\eta \rangle,
\eeq
which are given by
\beq
\label{del3}
  \langle {\cal Z}|  H_1^{OD} |{\cal Z}' \rangle  =  
      {1 \over 4 J} \left({ R^2 \over J}\right)^2  n^2 \sqrt{N^i_n (N^i_n-1) N^j_n{}'(N^j_n{}'-1)}.
\eeq
This can be computed similarly to (\ref{del2a}).
One should just multiply each term in (\ref{iijjod})--(\ref{cr5})
by 
\beq
\label{cf}
  {J^\N \over \sqrt{\Ot \, \Ot'}} {N^i_n (N^i_n-1) \over 2} {N^j_n{}' (N^j_n{}'-1) \over 2}
                           (N^i_n-2)!  \, (N^j_n{}'-2)!.
\eeq
The ingredients in (\ref{cf})  correspond to the normalization, the number of possible
choices of a pair out of  $N^i_n$ ($N^j_n{}'$) $\phi^i_n$'s ( $\phi^j_n$'s),
and the number of permutations of the leftover  $\phi^i_n$'s ( $\phi^j_n$'s).
Substituting $\Ot \approx \sqrt{J^\N N^i_n! N^j_n! \ldots}$ and
$\Ot' \approx \sqrt{ J^\N N^i_n{}'! N^j_n{}'! \ldots}$ into (\ref{cf}) one recovers
correct combinatorial factor in (\ref{del3}).

The expressions for diagonal matrix elements (\ref{tfdiag}) and (\ref{tfdiag2})
do not change when we allow $N^i_n>1$.
However (\ref{funnyc}) changes to
\beq
\label{funnyc2}
  {1 \over 2 J} \left( {R^2 \over J} \right)^2 \sum_{i,n} n^2 (N^i_n)^2.
\eeq
There is an additional contribution to the diagonal
matrix element, which is similar to (\ref{tfdiag2})
but with $n_l=n_k$.
To compute it, one has to follow the logic which led to 
(\ref{tfdiag2}) paying special attention to combinatorial factors.
We now have 
\beq
\label{intfree6}
 \delta \langle \Or_*(x) {\bar \Or}_*(0) \rangle_{g^0}= 
  {1 \over \Omega} \sum_{i,n} \sum
    \III{{\check \phi}^i_n}{{\check \phi}^i_{n}}{{\check \phi}^i_n}{{\check \phi}^i_n}
                    { \phi^{i_k}_{n_k}}{ \phi^{i_k}_{n_k}}=
     \sum_{i,n} {N^i_n (N^i_n-1) \over 4 J}
\eeq    
and
\beq
\label{intfree7}
 \delta \langle {\tilde \Or}(x) {\bar {\tilde \Or}}(0) \rangle_{g^0}= 0.
\eeq
since the diagram analogous to (\ref{intfree4}) with $m=n$ have been already
taken care of, and absorbed in the normalization constant.
The analog of (\ref{cr6}) is
\beq
\label{cr10}
   \delta \langle \Or_*(x) {\bar \Or}_*(0) \rangle_{g^1}{=} 
    {\gamma \over J}  \sum_{i,n} {N^i_n (N^i_n{-}1) \over 4} \bigg[
                                -\sum_k (q_{n_k}{+}q_{n_k}^*{-}2) 
    {+}2 (q_{n}{+}q_{n}^*{-}2)
          {-}{1 \over 2} (q_{2n}+q_{2n}^*) {+} 3  \bigg],
\eeq
while the contribution similar to (\ref{cr7}) is absent.
The analog of (\ref{cr8}) is
\beq
\label{cr11}
  \delta_3\langle {\tilde \Or}(x) {\bar {\tilde \Or}}(0) \rangle_{g^1}{=} 
  { \gamma \over J} \sum_{i,n} {N^i_n (N^i_n-1) \over 4} 
     \left[ 4 (q_{n}+q_{n}^*-2)+2  \right].
\eeq   
Finally, there is an analog of (\ref{cr9}) given by
\beq
\label{cr12}
   - \delta \langle \Or_*(x) {\bar{\tilde  \Or'}}(0){+} {\tilde \Or}(x) {\bar \Or}_*(0)\rangle_{g^1} =
- { \gamma \over  J} \sum_{i,n} {N^i_n (N^i_n-1) \over 2} (q_{n}+q_{n}^*).
\eeq
Combining (\ref{intfree6})--(\ref{cr12}) we get the following contribution
to the diagonal matrix element from the $\phi^i_n/\phi^i_n$ interactions
\beq
    {\beta \over J} \sum_{i,n}{ N^i_n (N^i_n-1) \over 4} \left( 4 q_n - {q_{2 n} \over 2} +c.c.\right) 
   =  - {1 \over J} \left( {R^2 \over J} \right)^2 \sum_{i,n} {n^2 N^i_n (N^i_n-1) \over 4}.
\eeq
Adding this to (\ref{funnyc2}) and then  replacing the last term
in (\ref{tfdiag}) with the resulting expression we recover the string
theory result (\ref{hdiagx}).
This concludes the matching of matrix elements between the
string and the gauge theory.


\end{document}